\documentclass[aps, twocolumn, pre, floatfix,  amsmath,amssymb,
aps,superscriptaddress,longbibliography]{revtex4-2}

\usepackage{amsmath,amssymb,amsthm}
\usepackage{bbold}
\usepackage{graphicx}

\newcommand{\llangle}{\ensuremath{\left\langle}}
\newcommand{\rrangle}{\ensuremath{\right\rangle}}
\newcommand{\ginoe}{\operatorname{GinOE}}

\newcommand{\R}{\mathbb{R}}
\newcommand{\one}{\mathbb{1}}

\renewcommand{\Im}{\operatorname{Im}}
\renewcommand{\Re}{\operatorname{Re}}

\newcommand{\tr}{\operatorname{tr}}

\graphicspath{{.}{figures/}}

\usepackage{color}

\definecolor{darkblue}{rgb}{0,0,.65}
\definecolor{darkgreen}{rgb}{0.3,0.6,0.3}
\definecolor{cyan1}{rgb}{0.0, 0.6, 0.6}
\usepackage[%
pdfstartview=FitH,%
breaklinks=true,%
bookmarks=true,%
colorlinks=true,%
anchorcolor=black,%
citecolor=darkgreen,
filecolor=black,%
menucolor=black,%
urlcolor=darkblue,%
linkcolor=blue,%
]{hyperref}

\usepackage{cleveref}  

\def\oper{{\mathchoice{\rm 1\mskip-4mu l}{\rm 1\mskip-4mu l}
{\rm 1\mskip-4.5mu l}{\rm 1\mskip-5mu l}}}

\begin{document}

\title{Random sparse generators of Markovian evolution and their spectral properties}

\begin{abstract}
The evolution of a complex multi-state system is often interpreted as a continuous-time Markovian process. To model the relaxation dynamics of such systems, we introduce an ensemble of random sparse matrices which can be used as generators of Markovian evolution. 
The sparsity is controlled by a parameter $\varphi$, which is the number of non-zero elements per row and column in the generator matrix.  Thus, a member of the ensemble is characterized by the Laplacian of a directed regular graph with $D$ vertices (number of system states) and $2 \varphi D$ edges with randomly distributed weights. 
We study the effects of sparsity on the spectrum of the generator.  Sparsity is shown to close the large spectral gap that is characteristic of non-sparse random generators.  
We show that the first moment of the eigenvalue distribution scales as $\sim \varphi$, while its variance is $\sim \sqrt{\varphi}$.
By using extreme value theory, we demonstrate how the shape of the spectral edges is determined by the tails of the corresponding weight distributions, and clarify the behavior of the  spectral gap as a function of $D$.  
Finally, we analyze complex spacing ratio statistics of ultra-sparse generators, $\varphi = \mathrm{const}$, and find that starting already at $\varphi \geqslant 2$, spectra of the generators exhibit universal properties typical of Ginibre’s Orthogonal Ensemble.
 
\end{abstract}

\newcommand{\dresdenTP}{Institut f\"{u}r Theoretische Physik, Technische Universit\"{a}t Dresden, D-01062 Dresden, Germany}

\author{Goran Nakerst}
\affiliation{\dresdenTP}

\author{Sergey Denisov}
\affiliation{NordSTAR - Nordic Center for Sustainable and Trustworthy AI Research, Pilestredet 52, N-0166, Oslo, Norway}
\affiliation{Department of Computer Science, Oslo Metropolitan University, N-0130 Oslo, Norway}

\author{Masudul Haque}
\affiliation{\dresdenTP}
\affiliation{Max-Planck-Institut f\"{u}r Physik komplexer Systeme, D-01187 Dresden, Germany}

\maketitle

\section{Introduction}

Continuous-time Markov chains (CTMCs) 
\cite{Anderson_book1991} provide a popular framework to model  dynamics of multi-state systems in diverse fields ranging from physics, chemistry, and biology 
\cite{Liggett_book1985interact_part_systems, Bharucha_1997Markov_processes, vanKampen_book2011_StochasticProcesses}
to economics~\cite{Lux_IFAC1995,Grasselli_Li_JournNetTheo2018} and game theory 
\cite{Hofbauer_Sigmund_1988dynamical_systems, Szabo_Gyorgy_2007evolutionary_games}.
CTMCs are used to model chemical reactions
\cite{McQuarrie_JourApplProb1967review_chem, Schnakenberg_RevModPhys1976, Pilling_AnnRevPhysChem2003, Gillespie_AnnRevPhysChem2007, Anderson_Kurtz_chapter2011chem, Lecca_2013review_chem, Rajvanshi_Venkatesh_EncSysBio2013},
gene regulation processes
\cite{Jong_JournCompBio2002, Hegland_etal_JourCompApplMath2007, Booth_Burden_ModSimScience2007, Lipan_MolLifeScience2014,  Sevier_etal_PNAS2016_transcrpt_noise},
quantum dynamics (approximated by rate equations)
\cite{Carroll_book1986ratee_quations_semiconductors, Vorberg_Wustmann_Ketzmerick_Eckardt_PRL2013, Prelovsek_Bonca_Mierzejewski_PRB2018,Mierzejewski_Prelovsek_Bonca_PRL2019, Liul_Shevchenko_LTP2023RateEquationQubits},
evolutionary game dynamics
\cite{Szabo_Gyorgy_2007evolutionary_games, Knebel_Weber_Krueger_NatComm2015, OvertonEtAl_JTheorBiology2019_StochasticEvolutionaryDynamics},
and many other processes. 
CTMCs are also the key element of such celebrated models of statistical physics as contact processes \cite{Harris_AnnProb1974, Marro_Dickman_1999Nonequilibrium_phase_transitions, Henkel_Hinrichsen_Haye_2008Nonequilibrium_phase_transitions}, 
zero-range processes \cite{Spitzer_AdvMath1970interacting_markov_process, Evans_Hanney_JPhysA2005}
and exclusion processes  like ASEP
\cite{%
    Spitzer_AdvMath1970interacting_markov_process,Liggett_book1985interact_part_systems, Spohn_1991Dynamics_interacting_particles, 
    Schuetz_Domany_JourStatPhys1993asep, Derrida_PhysRep1998ASEP, Liggett_1999stochastic_interacting_systems, Schuetz_PhaseTransCritPhen2001, 
    Golinelli_Mallick_JourPhysA2006ASEP, Chou_Mallick_Zia_RepProgPhys2011}.  
In some fields, CTMCs are known under the names `classical Markovian master equations' or `rate equations'.

A continuous-time Markovian evolution in finite discrete space consisting of $D$ states can be specified with a transition rate matrix $\cal{K}$~\cite{Anderson_book1991}, which is a generator of Markovian evolution. (It is  called `Kolmogorov operator' in Ref.~\cite{Tarnowski_Denisov_etal_PRE2021}). The  equation governing the evolution of a probability vector $P(t)$, defined on the state space,
\begin{equation}\label{eq:master_equation}
	\frac{d}{dt} P(t) = \mathcal{K} P(t),
\end{equation}
has the formal solution, $P(t) = \exp(t \mathcal{K})P_0$, where $P_0 = P(0)$ is the initial probability vector. The evolution of $P(t)$ is thus fully determined by the generator $\cal{K}$, especially by its spectral properties.  The fact that the operator $\mathcal{M}_t = \exp(t \cal{K})$ should map a non-negative vector onto another non-negative vector while preserving $\ell_1$-norm, means that $\cal{K}$ satisfies a set of constraints and these constraints have an effect on its spectral properties~\cite{Anderson_book1991}.
%which makes the latter the central object of study of this paper. 

In order to model the evolution of a complex system with CTMCs, we would have to first design a specific Kolmogorov operator. Taking into account the large variety of existing models, it would be beneficial to figure out universal properties of $\cal{K}$-generators, i.e., properties that are typical rather than specific to a particular model. The first step in this direction is to define {\it random} ensembles of generators. A similar situation arises in the case of unitary time-continuous evolution, where the corresponding generators (quantum Hamiltonians)  were explored and classified by using the powerful toolbox  of random matrix theory (RMT) \cite{Wigner_1956conference, Wigner_1957, Dyson_JourMathPhys1962rmt1, Dyson_JourMathPhys1962rmt2, Dyson_JourMathPhys1962rmt3}.  Implementation of this idea resulted in the creation of Quantum Chaos theory \cite{Haake_1991chaosbook, Stockmann_1999, Braun_2001dissipativechaosbook} which made - and is still making -- a strong impact on many-body quantum physics, both theoretical \cite{Borgonovi_et_al_PhysRep2016chaos_therm} and experimental (see, e.g., Ref.~ \cite{Roushan_et_al_Science2017localization_qubits}).

Recently,  RMT-based approaches were developed to analyze spectral properties of random generators of open quantum (Lindblad operators)~\cite{Denisov_et_al_PRL2019, Can_JourPhysA2019randLindblad, Lange_Timm_Chaos2021Lindblad, Sa_Ribeiro_Prosen_JournPhysA2021random_Liouvillian} and  classical  (Kolmogorov operators)~\cite{Timm_PRE2009, Bordenave_Caputo_Chafai_CommPureApplMath2014sparse_Markov, Tarnowski_Denisov_etal_PRE2021} Markovian evolution. The considered generators, both quantum and classical, were on purpose constructed in a completely random way  - up to the constraints that make them legitimate generators. In the case of Kolmogorov operators, this means that they are represented by dense matrices~\cite{Timm_PRE2009,Tarnowski_Denisov_etal_PRE2021}. It was shown that the spectral density of such operators represents a free sum of a uniform disc and a Gaussian distribution which results in a distinctive spindle-like shape~\cite{Tarnowski_Denisov_etal_PRE2021}, as shown in  Figure~\ref{fig:spectra_random} (a). This density is universal, in the sense that the particular way the operators are sampled does not affect the shape of the spindle (but may affect its position on the real axis and its overall scaling)~\cite{Tarnowski_Denisov_etal_PRE2021}.

In contrast to the random Kolmogorov operators, for most applications and known models, the  corresponding $\cal{K}$-generators  are represented by {\it sparse} matrices.  This is a consequence of locality and other topological constraints imposed on the allowed transitions in the state spaces of the models.  For many-component or many-particle systems, elements of the generator matrix typically represent changing multiple (or all) components of the system simultaneously, e.g., for an exclusion process, a generic matrix element could represent correlated hopping of many particles.  Since such processes are usually absent in physically (biologically, economically,...)-motivated models, most elements of the corresponding $\cal{K}$-matrices are zeros.  

\begin{figure*}[t]
\begin{center}
	\includegraphics[width=\textwidth]{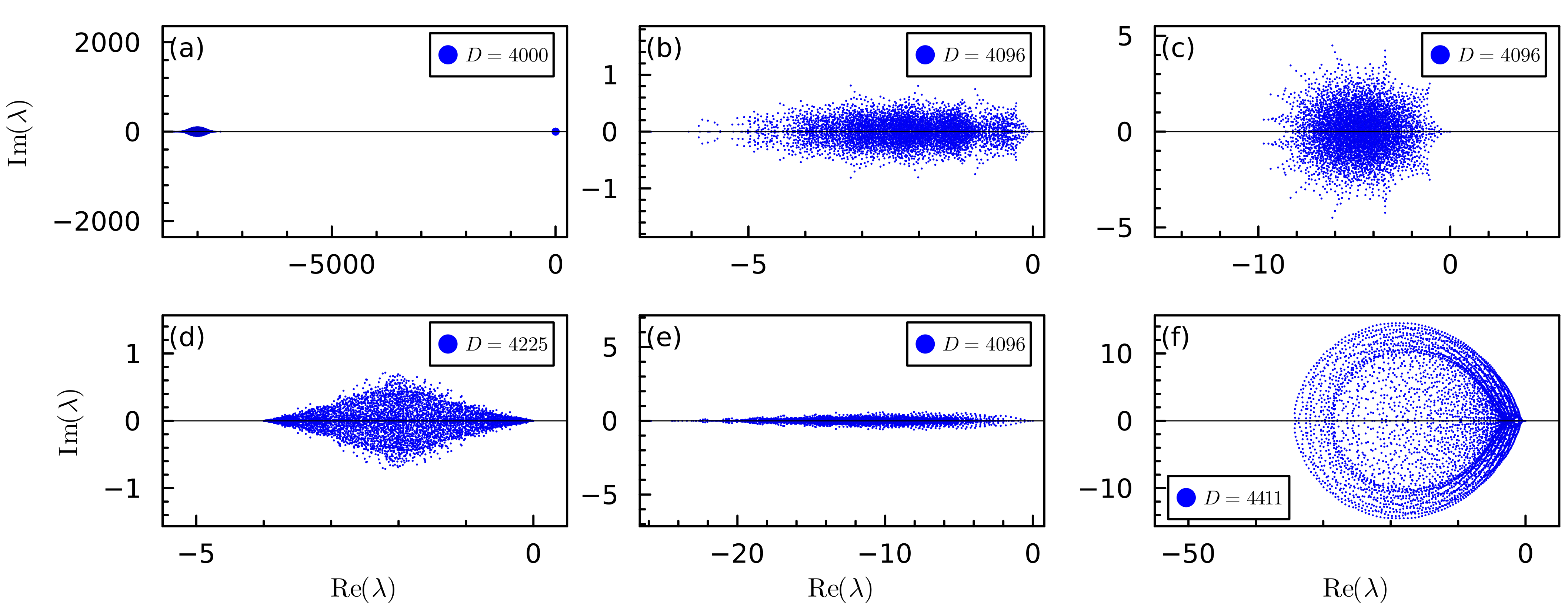}
	\caption{Spectra of  generators of Markovian evolution, Eq.~\eqref{eq:master_equation}. \textbf{(a)} Dense (non-sparse) random generator with $\chi_2^2$ edge weight distribution, \textbf{(b)} totally asymmetric simple exclusion process (TASEP) on a ring with staggered hopping probabilities~ \cite{Sa_Ribeiro_Prosen_PRX2020}, \textbf{(c)} asymmetric simple exclusion process {ASEP} on a chain with open boundary conditions and next nearest neighbor terms, \textbf{(d)} a process of particle hopping on an open boundary grid with random hopping probabilities, \textbf{(e)} a contact process on a chain \cite{Harris_AnnProb1974}, \textbf{(f)} a gene transcription model from Ref.~\cite{Sevier_etal_PNAS2016_transcrpt_noise}. In each plot the real and imaginary axes have the same scale.  The models are described in Appendix~\ref{sec:appendix_systems_description}. \label{fig:spectra_systems}}
\end{center}
\end{figure*}

Sparsity affects the spectra, $\{\lambda_i\}$, $i=1,2,...,D$, of the corresponding generators. Most noticeably, the spectral gap, i.e., the distance between  $\lambda_1=0$
and the eigenvalue closest to it, $\gamma_* = \min \{| \Re \lambda_i|\}$, does not grow with the increase of number of states $D$. 
This is in sharp contrast to the case of dense random generators~~\cite{Denisov_et_al_PRL2019, Lange_Timm_Chaos2021Lindblad, Sa_Ribeiro_Prosen_JournPhysA2021random_Liouvillian} .  
In Figure~\ref{fig:spectra_systems} we contrast the case of dense random stochastic matrices (a) with various model systems described by sparse Markov generators (b-f).  

The large gap of dense random generators implies that
even the slowest decaying mode of a generic initial probability vector converges rapidly to the equilibrium state, the relaxation time (inverse of the spectral gap) decreasing inversely in the state space size $D$.  In contrast, physical generators
of CTMCs in general have spectral gaps and relaxation
times functionally depending on $D$ very differently than
(anti-)linearly; see, e.g. \cite{Gwa_Spohn_PRA1992bethe, Kim_PRE1995bethe,Golinelli_Mallick_JournPhysA2004,Golinelli_Mallick_JournPhysA2005} for the example of the exclusion process.   This difference of behavior suggests that it is more suitable to model physical CTMC generators by sparse rather than dense random matrices.

Our motivation is to refine the RMT approach to random Kolmogorov operators by including sparsity which is characteristic to physically relevant  $\cal{K}$-generators.  We specify an ensemble  of random matrices of fixed  sparsity $\varphi$ as an ensemble of  negative combinatorial Laplacians of random regular directed graphs. The sparsity is controlled by the vertex degree $\varphi$ which is equal to  the number of non-zero elements per row and column of the generator matrix.  In graph terms, this means that each vertex has exactly $\varphi$ incoming and $\varphi$ outgoing edges.  The non-zero elements (edge weights) are taken to be random, positive, independent, and identically distributed (iid). 

A similar setup was studied in Ref.~\cite{Bordenave_Caputo_Chafai_CommPureApplMath2014sparse_Markov}, where an ensemble of oriented Erd\H{o}s-R\'{e}nyi graphs~\cite{Bollobas_2001random_graphs}, parameterized with edge probability distribution $p(D)$, was used.  The vertex distribution in this case is binomial-distributed~\cite{Bollobas_2001random_graphs}, and not constant as in our case. However, one might expect similar behavior in the $D\to\infty$ limit with the correspondence $p(D)=\varphi/D$. The authors of Ref.~\cite{Bordenave_Caputo_Chafai_CommPureApplMath2014sparse_Markov} considered the regime $Dp(D) \gg (\log D)^6$, which they found to have the same universal properties as in the non-sparse case.  In this work, we consider sparsity beyond this limit, including specifically $\varphi\sim D^0$ (vertex degree not growing with $D$) and $\varphi\sim\log D$.     

In this paper, we investigate the dependence of spectral properties of the sparse Kolmogorov operators on  sparsity parameter $\varphi$, number of states (dimension of the state space) $D$, and on the edge weight distribution, i.e., on the distribution of the nonzero  elements of matrix $\mathcal{K}$.  
Explicit results are mostly derived for  $\chi_2^2$  and uniform weight distributions; however, these results can be  adapted to other weight distributions.  

We consider both the bulk of the spectral distribution and its edges.

As for the bulk, we focus on its position $\mu$ (the mean of the corresponding eigenvalue distribution) and its variances along the real and imaginary axes (standard deviations of the distribution of the real and imaginary eigenvalue parts, respectively). The first variance estimates the spread of the relaxation rates, while the second one gives the timescales of the oscillations during the relaxation.

As for the edges,  we address  the spectral gap and the extent of the spectrum along the real axis (the real part of the eigenvalue with largest absolute real part).  These determine respectively the slowest and fastest time scales of relaxation to the steady state.  The spectral gap is of physical interest for multi-state Markov processes, see, e.g., \cite{Gwa_Spohn_PRA1992bethe, Kim_PRE1995bethe,Golinelli_Mallick_JournPhysA2004, Golinelli_Mallick_JournPhysA2005, deGier_Essler_JSM2006,deGier_Essler_JPhysA2008, deGier_Finn_Sorrell_JPhysA2011, Prolhac_JPhysA2014, Prolhac_JPhysA2017} for the ASEP and \cite{Henkel_Schollwoeck_JPhysA2001, Hooyberghs_Igloi_Vanderzande_PRE2004} for the contact process.  
The horizontal extent, in addition to its interpretation as the fastest timescale for CTMCs, is also relevant in the graph theory interpretation, e.g., to quantify the computational complexity of the community detection problem~\cite{Abbe_et_al_IEEETranNet2014, Bandeira_FoundCompMath2018} and the max-cut  \cite{Delorme_Poljak_MathProg1993max_cut, Goemans_Williamson_JACM1995max_cut} problems.

We demonstrate that the position and variance of the spectral bulk of sparse Kolmogorov operators scale as $\sim\varphi$ and $\sim\sqrt{\varphi}$, respectively. These characteristics do not  depend on $D$ but on the first and second moments of the weight distribution. The dependence of the spectral edges on the weight distribution is less straightforward and highly non-universal. In particular, we show that,  in the regime of high sparsity, $\varphi \ll D$, the spectral gap (horizontal extent) depends only on the left (right) tail of the weight distribution. We evaluate the dependence of the spectral gap on $\varphi$ and $D$ for weight distributions with exponential and  power-law tails.

We consider the cases of $\chi_2^2$ and uniform weight distributions in detail. For these distributions, we find that the spectral gap and the horizontal extent of the spectrum can be  approximated by the largest and smallest diagonal entry of the generator matrix, respectively. 
Using the conjecture that this correspondence holds in general, we use extreme value theory (EVT) \cite{deHaan_Ferreira_2006EVT,Embrechts_Kluppelberg_Mikosch_1997ExtremeEvents} to analytically derive dependencies of the spectral edges on $\varphi$ and $D$.  In particular, we infer that the distributions of spectral edges only depend on the tails of the weight distributions.

Finally, we analyze correlations of the eigenvalues of sparse Kolmogorov operators. We show that, for $\varphi\ge 2$, the complex spacing ratio distributions~\cite{Sa_Ribeiro_Prosen_PRX2020} of the spectral bulks follow the  distribution typical to Ginibre’s Orthogonal Ensemble. 

The paper is organized as follows. In Section~\ref{sec:sparse_random} we introduce an ensemble of sparse random Kolmogorov operators. We analyze the  bulk of the spectral distributions of the ensembles in Section~\ref{sec:bulk}. In Sections~\ref{sec:gap} and \ref{sec:extent} we address the spectral gap and the horizontal extent of the spectrum, respectively. A discussion on correlations between eigenvalues in terms of the complex spacing ratio follows in Section~\ref{sec:ratios}.  We conclude with a summary of our results  in Section~\ref{sec:discussion}. Appendices contain information on the models whose spectra are presented in Figure~\ref{fig:spectra_systems}, details of the sampling of sparse random Kolmogorov operators, and the details of analytical derivations.

\section{Random sparse Kolmogorov operators}\label{sec:sparse_random}

In this section, we first recall some basic properties of Kolmogorov operators and review the case of full random $\mathcal{K}$-matrices.  We then define an ensemble of random sparse operators. In what follows, matrices will be referred to by calligraphic letters (e.g., $\mathcal{K}$) while their elements will be denoted by non-calligraphic letters (e.g., $K_{ij}$).

\subsection{Basic information}\label{sec:preliminaries}

In order to be qualified as a Kolmogorov operator, a matrix $\cal{K}$ has to fulfill 
two conditions, (i) all its off-diagonal elements have to be real and non-negative,  $K_{ij} \geq 0$, $i\neq j$, and (ii) the sum over every column should be zero. The latter is fulfilled by setting all the diagonal elements as
\begin{equation}\label{eq:cons_prob}
	K_{ii} = - \sum_{j\neq i} K_{ji}.
\end{equation}
The first condition guarantees the preservation of the non-negativity of a vector during the evolution induced by Eq.~\eqref{eq:master_equation}, while the second one guarantees the preservation of the $\ell_1$-norm of the vector.

The spectrum of $\cal{K}$ is in general complex. Since $\cal{K}$ maps real vectors onto real vectors, the spectrum is invariant under complex conjugation, so all complex eigenvalues come in conjugated pairs. The spectrum contains at least one eigenvalue $\lambda_1=0$ with right eigenvector corresponding to the steady  state. By  virtue of the Perron-Frobenius theorem \cite{Perron_1907, Frobenius_1912, Keizer_JournStatPhys1972}, the components of the steady state vector can be chosen to be non-negative, which makes it, after normalization, a probability vector.
Any Kolmogorov operator can be represented in terms of a real non-negative matrix, $\cal{M}$, $M_{ij} \geq 0$,
\begin{equation}
	\cal{K} = \cal{M} - \cal{J},
\end{equation}
where elements of the diagonal matrix $\cal{J}$ are $J_{jj} = \sum_i M_{ij}$. 

We now briefly review the case of dense (non-sparse) random Kolmogorov operators~\cite{Timm_PRE2009, Bordenave_Caputo_Chafai_CommPureApplMath2014sparse_Markov, Tarnowski_Denisov_etal_PRE2021}.  Elements $M_{ij}>0$ are i.i.d. sampled from a distribution with density $p(x)$ and first two moments $\mu_0=\int xp(x)dx$ and $\sigma^2_0 = \int (x-\mu_0)^2p(x)dx$. The particular choice of distribution does not  play an essential role (provided that it is not very pathological).
For example, we could sample a matrix $\cal{Z}$ from Ginibre's Unitary Ensemble (GinUE) and  then square its elements, $M_{ij} = |Z_{ij}|^2$~\cite{Tarnowski_Denisov_etal_PRE2021}).
The matrix $\cal{M}$ is then  full in the sense that, with probability $1$, all its elements are different from zero.

The elements of the matrix $\mathcal{M}$ are i.i.d., thus, in the asymptotic limit, its spectral density is a uniform disk of radius $\sqrt{D}\sigma_0$, with the center at $0$. 
In the dense limit, the elements of $\cal{J}$ are sums of $D$ independent random variables, so its elements can be approximated with Gaussian-distributed random variables with mean $D\mu_0$ and variance $D\sigma_0^2$.

Following the RMT approach~\cite{Tarnowski_Denisov_etal_PRE2021}, the Kolmogorov operator in Eq.~\eqref{eq:cons_prob} can be modelled as
\begin{equation}\label{eq:RMT_model}
{\cal K}_{\mathrm RM} = -\mu_0 D \cdot \oper+ \sigma_0 \sqrt{D}({\cal G} - {\cal D}),
\end{equation}
where $\cal{G}$ is a member of Ginibre's Orthogonal Ensemble (GinOE) and $\mathcal{D}$ is a diagonal matrix. Elements of $\mathcal{G}$ and $\mathcal{D}$ are sampled from the normal distribution of zero mean and unit variance. Here $\sigma_0 \sqrt{D}\cdot{\cal G}$ models $\cal{M}$ while $\cal{J}$ is approximated as $\mu_0 D \cdot \oper + \sigma_0 \sqrt{D}\cdot{\cal D}$.

\begin{figure*}
	\begin{center}
		\includegraphics[width=\textwidth]{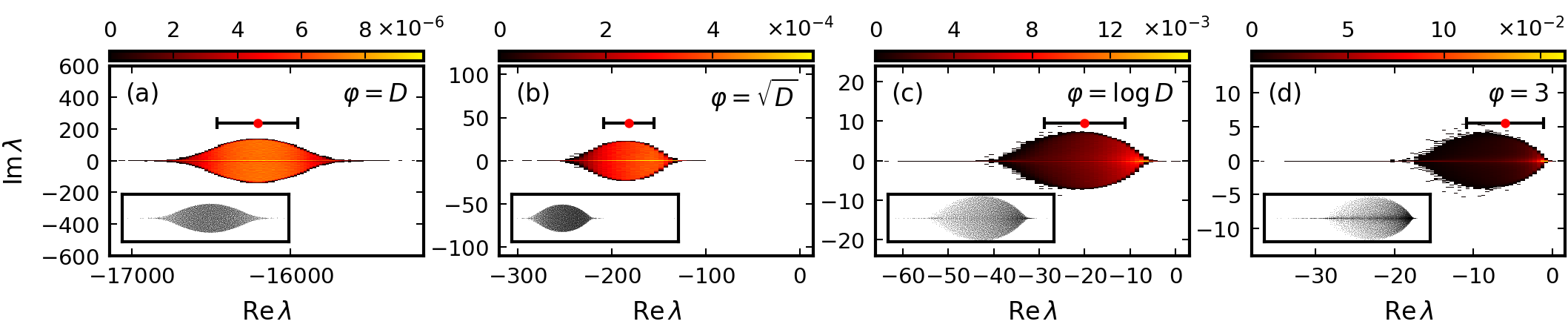}
		\caption{\label{fig:spectra_random} Spectral densities of random Kolmogorov operators with $\chi_2^2$ weight distribution. The matrix size is $D\approx 8000$ and the densities are estimated with $100$ samples. White areas contain no eigenvalues.  \textbf{(a)} Dense matrix without the zero eigenvalue, \textbf{(b)} sparse matrix with $\varphi=\sqrt{D}$ non-zero elements per row and column, \textbf{(c)} $\varphi=\log D$ and \textbf{(d)} $\varphi=3$. The insets show spectra of  single realizations.  In each plot, the real and imaginary axes have the same scale.  The red dots mark the location of $\mu(\lambda)$, given by Eq.~\eqref{eq:est_mean_def}, and the intervals shown in black are $[\mu(\lambda)-\sigma(\lambda),\mu(\lambda)+\sigma(\lambda)]$, where $\sigma(\lambda)$ is given by Eq.~\eqref{eq:sigma2_re_def}. %Both are shifted in the imaginary axis for visualization purpose.
  }
	\end{center}
\end{figure*}

The spectral density of the non-trivial part, ${\cal K}'={\cal G}+{\cal D}$, is a free convolution of a disk and a Gaussian distribution along the real axis, which results in a spindle-like shape. Figure~\ref{fig:spectra_random}~(a) presents both the spectrum of a single random  dense Kolmogorov operator and histogram obtained with $100$ samples. 

Alternatively, we can state that the spectral density of the rescaled generator 
\begin{equation}\label{eq:scaling_dense}
{\cal K}' = \frac{1}{\sigma \sqrt{D}}({\cal K} + \mu_0 D \cdot\oper)
\end{equation}
is expected, in the asymptotic limit, to be the $D$-independent spindle ("an additive Gaussian deformation of the circular law", according to Ref.~\cite{Bordenave_Caputo_Chafai_CommPureApplMath2014sparse_Markov}).

The spectrum of the random non-sparse generator has a large gap which scales as $D$, as seen in Figures \ref{fig:spectra_systems} and \ref{fig:spectra_random}.  We will see that this feature is strongly affected when we introduce sparsity.

\subsection{Ensemble of sparse random Kolmogorov operators as a set of oriented graphs}

The operator ${\cal K}$ described in the introduction can be considered as the negative Laplacian of a random directed graph with positive, iid edge weights, without self-loops, and with fixed vertex degree equal to $\varphi$. 

For example, the graph corresponding to the ${\cal K}$ generator of a process of a particle hopping on a $d$-dimensional hypercubic lattice with periodic boundary conditions and random hopping rates is a  particular (to the nearest-neighbor connections) realization of the ensemble with $\varphi=2d$. Figure~\ref{fig:spectra_systems} (d) shows an example spectrum for $d=2$.

The regularity of the graphs ensures that, with probability $1-O(D^{-\varphi-1})$, they are  strongly connected as long as $\varphi \ge 2$ \cite{Pittel_2018}. Strong connectivity is a good feature because it means that the matrix ${\cal K}$ is not of block-diagonal structure and the state space  is not partitioned into disconnected subsets. 
As there is only one strongly connected component, there is only one absorbing component. This implies that the multiplicity of the zero eigenvalue is one and the steady state is unique. Finally, every state in the state space is reachable from  every other state. The steady state, therefore, has all states populated. 

Some physical models motivating this study presented in Figure~\ref{fig:spectra_systems} are - except for the contact process - all strongly connected. The contact process is only effectively strongly connected. It has two strongly connected components, where one is a single vertex and the other includes the remaining $D-1$ vertices. The giant component is the only absorbing component and consequently, the steady state is unique.

We will focus on sparse generators with $\varphi\ge 2$ and will discuss $\varphi=1$ in Section \ref{sec:ratios}. 

The physical models presented in Figure~\ref{fig:spectra_systems} motivate us to focus on two types of dependencies of $\varphi$ on the matrix size $D$, namely $\varphi = \mathrm{const}$  and $\varphi\sim\log D$. For generators of single particle hopping models - an example is shown in Figure~\ref{fig:spectra_systems} (d) - the average number of non-zero elements per column and row is constant and independent of $D$. It increases logarithmically with $D$ in many-body hopping models such as the ASEP or the contact process, Figure~\ref{fig:spectra_systems} (b), (c), and (e). There is no simple dependence of $\varphi$ on $D$ in the gene transcription model, Figure~\ref{fig:spectra_systems} (f), as the matrix size $D$ is controlled by multiple parameters, see Appendix~\ref{sec:appendix_systems_description}.

What can we say about spectral densities of the ultra-sparse $\mathcal{K}$-generators, with $\varphi = \mathrm{const}$? A 'naive' adjustment of the RMT approach, which consists in describing  the elements of a sparse matrix $\cal{M}$  with probability density function $\widetilde{p}(x) = (1 - \frac{\varphi}{D})\delta(x) + \frac{\varphi}{D} p(x)$, re-scaling the mean and variance accordingly, and then using the RMT model, Eq.~\eqref{eq:RMT_model}, would not work here for two reasons. First, the spectral densities of such sparse matrices cannot be approximated with members of 'dense' RMT ensembles. Second, the Central Limit Theorem no longer applies and the entries of matrix $\mathcal{J}$ cannot be approximated with normal random variables (the entries become distribution-specific).

\section{Position and width of the  bulk of the spectrum}\label{sec:bulk}

In this section, we analyze the dependence of the position and horizontal width of the bulk of the spectrum on the sparsity parameter $\varphi$ and the matrix dimension $D$. 
We first provide (Subsections~\ref{sec:bulk_location} and \ref{sec:bulk_width})  expressions and bounds for the position and the width, characterized respectively by the mean $\mu(\lambda)$ of all eigenvalues and the standard deviation $\sigma(\Re\lambda)$ of the real parts of the eigenvalues.  These results are expressed in terms of the 
 mean and standard deviation of the weight distribution (distribution of non-zero elements of the Kolmogorov operator  ${\cal K}$), denoted by $\mu_0$ and  $\sigma_0$ respectively.

Since the most prominent effect of sparsity is to reduce the parametrically large gap seen in the full random case, it is instructive to analyze the ratio $\alpha = |\mu(\lambda)|/\sigma(\Re\lambda)$.  This quantity provides insight into the distance of the bulk of the spectrum from the origin, relative to the size of the bulk.  Subsection~\ref{sec:alpha} is devoted to an analysis of the ratio $\alpha$.

Numerical results presented in this section are obtained by sampling edge weights from the $\chi_2^2$ and the standard uniform distribution.

The spectrum of dense generators ($\varphi=D-1$) consists of two distinct parts - an eigenvalue $\lambda_1 = 0$ and the rest of the eigenvalues forming the spectral bulk away from the imaginary axis, as shown in Figure~\ref{fig:spectra_systems} (a) and Figure~\ref{fig:spectra_random} (a).  In contrast, the bulk of the spectrum is much closer to the imaginary axis for $\varphi \ll D$, as seen in Figure~\ref{fig:spectra_random} for (b) $\varphi=\sqrt{D}$, (c) for $\varphi=\log D$ and (d) for $\varphi=3$.  For $\varphi=\sqrt{D}$, the bulk of the spectrum  is visibly separated from the zero, as in the dense case.  In fact, the spectral boundary is given by the same spindle (properly rescaled). Whether the spectral distribution is separated from zero for $\varphi=\log D$ and $\phi=3$ is difficult to say  with certainty from the available numerical data ($D\approx 8000$).

\subsection{Position}\label{sec:bulk_location}
The position of the spectral bulk of ${\cal K}$ can be identified with the mean $\mu(\lambda)$ of eigenvalues $\lambda_i$,
\begin{equation}\label{eq:est_mean_def}
	\mu(\lambda) = \left\langle \frac{1}{D} \sum_{i=1}^D \lambda_i \right\rangle,
\end{equation}
where the average $\langle\dots\rangle$ is taken over the ensemble of random Kolmogorov operators  described in Section~\ref{sec:sparse_random}. Because the eigenvalues  are either real or come in complex conjugate pairs, the mean of the spectral bulk is real, $\mu(\lambda)= \mu(\Re \lambda)$.

A simple calculation, presented in Appendix~\ref{sec:appendix_bulk}, shows that $\mu(\lambda)$ can be expressed as
\begin{equation}\label{eq:est_mean}
	\mu(\lambda) = \left\langle \frac{1}{D}\tr(\mathcal{K}) \right \rangle = - \varphi \mu_0,
	%&= -(D-1)\mu \approx - D \epsilon \mu_0 = -\varphi \mu_0 \\
	%\sigma_\lambda &= \sqrt{\frac{D-1}{D}} \sqrt{\varphi}\sqrt{(\sigma_0^2 + \mu_0^2)} \approx \sqrt{\varphi}\sqrt{(\sigma_0^2 + \mu_0^2)}.
\end{equation}
The averaging $\langle\dots\rangle$ over the matrix ensemble in Eq.~\eqref{eq:est_mean_def} and Eq.~\eqref{eq:est_mean} is, in principle, not needed since typicality is expected, i.e., for large enough $D$, a single sample will display all the spectral features of the ensemble. This is because the quantity $\frac{1}{D}\tr(\mathcal{K})$ is concentrated around its average $\llangle \frac{1}{D} \tr(\mathcal{K}) \rrangle$ for increasing $D$, as shown in Appendix~\ref{sec:appendix_bulk}.

For the four different dependencies of $\varphi$ on $D$ shown in Figure~\ref{fig:spectra_random},
Eq.~\eqref{eq:est_mean} implies the following: For $\varphi = \mathrm{const}$, the  mean is independent of the matrix size $D$. For $\varphi=\log D$ ($\varphi=\sqrt{D}$) the mean decreases logarithmically with $D$ (as $\sim \sqrt{D}$) and for $\varphi = D$ the mean decreases linearly with $D$ as is expected for the dense generators~\cite{Timm_PRE2009}.

In Figure~\ref{fig:spectra_random},  the location $\mu(\lambda)$ of generator matrices $\cal{K}$ is indicated with a red dot in each panel. The real part of the dot resides in the bulk of the spectrum for every dependence of $\varphi$ on $D$ shown in Figure~ \ref{fig:spectra_random}.

\subsection{Horizontal width}\label{sec:bulk_width}

In Section~\ref{sec:bulk_location} we investigated where the bulk of the spectrum is located in the complex plane.  We now analyze the width of the distribution.  We are especially interested in the horizontal width.

We characterize the width of the bulk spectrum, both in the real and imaginary directions, $\Re \lambda$ and $\Im \lambda$, using the estimated variances 
\begin{align}
	\sigma^2(\Re\lambda) &= \left\langle \frac{1}{D} \sum_{i=1}^D \left(\Re\lambda_i - \frac{1}{D}\sum_{j=1}^D \lambda_j \right)^2 \right\rangle \label{eq:sigma2_re_def}\\
	\sigma^2(\Im\lambda) &= \left\langle \frac{1}{D} \sum_{i=1}^D (\Im\lambda_i)^2 \right\rangle,
\end{align}
where we used the fact that $\sum_{j=1}^D \lambda_j$ is real. 

Because the eigenvalues appear in complex conjugate pairs,   $\sigma^2(\Re \lambda)$ and $\sigma^2(\Im \lambda)$ are related to the estimated complex pseudo-variance via
\begin{align}
	\sigma^2(\lambda) 
	&= \left\langle \frac{1}{D} \sum_{i=1}^D \left(\lambda_i - \frac{1}{D}\sum_{j=1}^D \lambda_j \right)^2 \right\rangle \nonumber \\
	&= \sigma^2(\Re\lambda) - \sigma^2(\Im\lambda).
\end{align}
The estimated pseudo variance lower bounds the estimated variance of the real parts of the eigenvalues, $\sigma^2(\lambda) \le \sigma^2(\Re \lambda)$.

The complex pseudo variance can be analytically calculated for the ensemble of random generator matrices as
\begin{align}\label{eq:sigma2_analytical}
	\sigma^2(\lambda) &= \llangle \frac{1}{D}\tr(\mathcal{K}^2) \rrangle - \llangle \frac{1}{D^2} \tr(\mathcal{K})^2 \rrangle \nonumber \\
	&= \varphi \left( \sigma_0^2 + \frac{\varphi}{D} \mu_0^2 - \frac{1}{D}\sigma_0^2 \right) .
\end{align}
Details of the calculation are provided in Appendix~\ref{sec:appendix_bulk}. The bound of the estimated real variance by the pseudo variance together with Eq.~\eqref{eq:sigma2_analytical} leads to the asymptotic lower bound of $\sigma(\Re \lambda)$ in terms of the sparsity parameter $\varphi$. As $1 \le\varphi \le D$, the estimated horizontal width of the bulk spectrum cannot grow asymptotically slower than $\sqrt{\varphi}$,
\begin{equation}\label{eq:sigma_bound}
	 \sigma(\Re \lambda) \gtrsim \sqrt{\varphi} .
\end{equation}
Numerically, we find that the bound in Eq.~\eqref{eq:sigma_bound} is asymptotically sharp for $\varphi\ll D$, as shown in Figure~\ref{fig:alpha} through the ratio $\alpha$ of mean $\mu(\Re \lambda)$ and width $\sigma(\Re \lambda)$. The collapse of the data points in Figure~\ref{fig:alpha} (c) implies that $\sigma(\Re \lambda) \sim \sqrt{\varphi}$. 

\subsection{Ratio of mean and horizontal width}\label{sec:alpha}

\begin{figure}
	\begin{center}
		\includegraphics[width=\columnwidth]{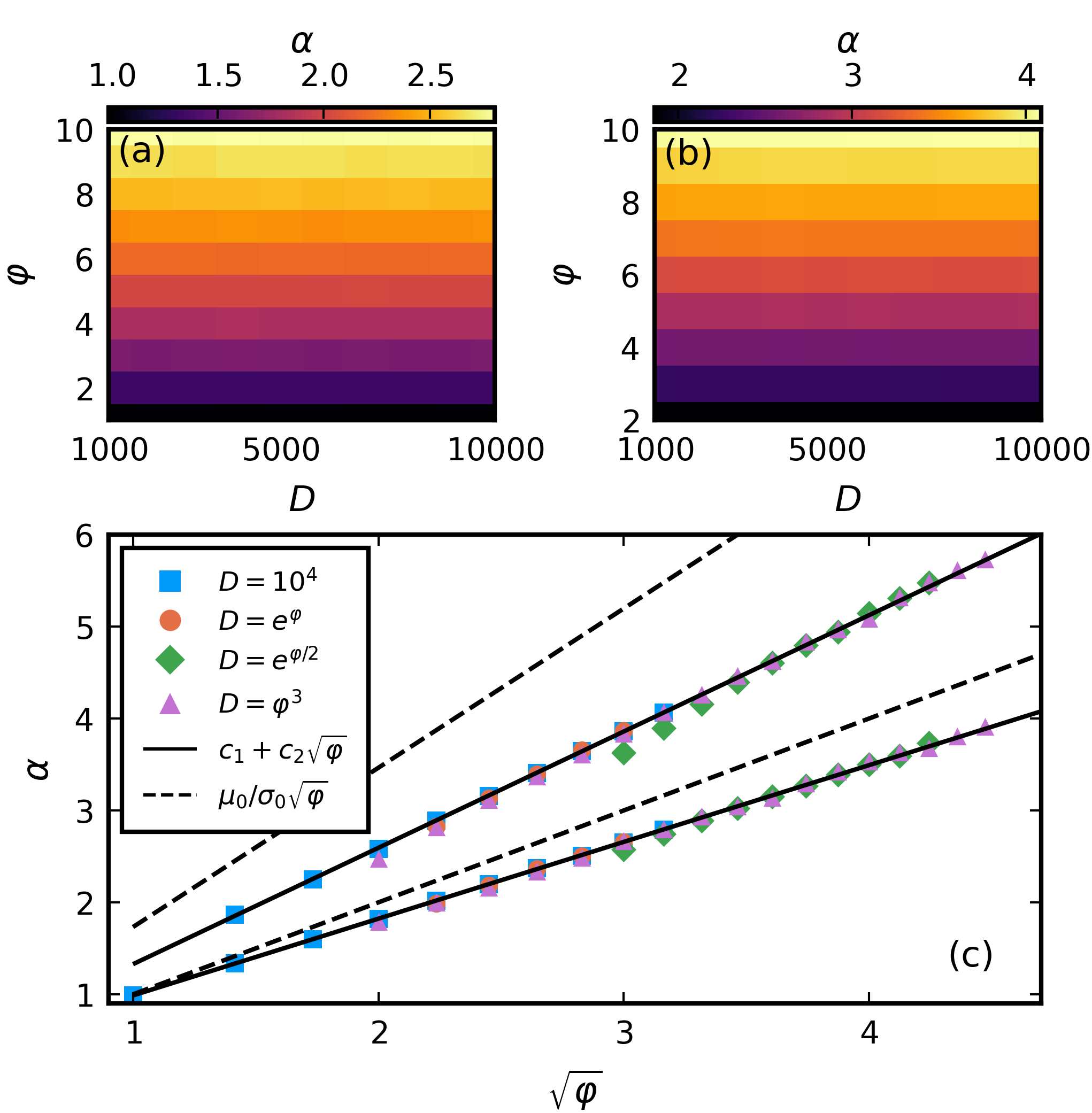}
		\caption{\label{fig:alpha} Ratio $\alpha$ of mean $\mu(\Re\lambda)$ and horizontal width $\sigma(\Re \lambda)$ of the bulk of the spectrum of sparse random Kolmogorov operators with \textbf{(a)} $\chi_2^2$ and \textbf{(b)} standard uniform weight distributions.  \textbf{(c)} $\alpha$ as a function of $\sqrt{\varphi}$. The bottom markers correspond to $\chi_2^2$ and the top to uniform distribution. Dependencies of $\varphi$ on $D$ are $\varphi\equiv$ constant, $\varphi=\log D$, $\varphi=2\log D$, and $\varphi=D^{1/3}$. The black solid lines correspond to $\alpha = c_1 + c_2 \sqrt{\varphi}$ ($c_{1,2}$ given in the main text) and the dashed lines denote $\alpha=\mu_0/\sigma_0 \sqrt{\varphi}$.
  }
	\end{center}
\end{figure}

In this section, we combine the information of the location of the spectrum given by Eq.~\eqref{eq:est_mean_def} and the horizontal width of the bulk given by Eq.~\eqref{eq:sigma2_re_def} into the ratio
\begin{equation}\label{eq:alpha_def}
	\alpha = \frac{|\mu(\Re \lambda)|}{\sigma(\Re\lambda)}.
\end{equation}
This quantifies how close the bulk spectrum is, relative to its size, to the stationary value $\lambda_1=0$. i.e., to the imaginary axis. For $\alpha = O(1)$ the estimated width of the bulk is of the same order as the estimated mean, thus the spectrum is located close to 0. For $\alpha \gg 1$ the estimated mean is much bigger than the horizontal width of the bulk and the bulk of the spectrum is far away from $0$. 

The analytical result for the estimated mean of the spectrum,  Eq.~\eqref{eq:est_mean}, together with the asymptotic bound on the standard deviation of the real parts of the spectrum, Eq.~\eqref{eq:sigma2_analytical}, imply the following asymptotic bound on $\alpha$
\begin{equation}\label{eq:alpha_bound_analytical}
	\alpha \lesssim \sqrt{\varphi}.
\end{equation}
Numerically, we observe that the bound in Eq.~\eqref{eq:alpha_bound_analytical} is asymptotically tight for $\varphi\ll D$, i.e.
\begin{equation}\label{eq:alpha_numerical}
	\alpha \approx c_1 + c_2 \sqrt{\varphi},
\end{equation}
for  constants $c_1$ and $c_2$.  Since $\mu(\lambda)$ scales linearly with $\varphi$, this behavior is consistent with $\sigma(\Re\lambda) \sim \sqrt{\varphi}$, stated previously.
The constants are found to be $c_1\approx 0.15$ ($\approx0.1$) and $c_2\approx 0.84$ ($\approx 1.3$) for the  $\chi_2^2$ (uniform) distribution.

Numerical results for $\alpha$ are summarized in Figure~\ref{fig:alpha}. For each combination of $\varphi$ and $D$, $\alpha$ is averaged over $n$ samples of random generators such that $nD = 50'000$. The weight distribution is the $\chi_2^2$ distribution in (a) and in the lower part of (c), and is the uniform distribution in $[0,1]$ in (b) and in the upper part of Figure~\ref{fig:alpha} (c). We have found that these results are qualitatively the same for exponentially distributed edge weights.

In Figure~\ref{fig:alpha} (a,b) we show the value of $\alpha$ as a function of $D$ and $\varphi$. On the $x$-axis $D$ varies in steps of $10^3$ between $10^3$ and $10^4$.  We observe that $\alpha$ increases with $\varphi$ and is independent of $D$,  as predicted by Eq.~\eqref{eq:alpha_numerical}. 
In Figure~\ref{fig:alpha} (c) we show $\alpha$ as a function of $\varphi$ for different dependencies of $\varphi$ on $D$. In all the cases, values of $\alpha$ collapse onto the black solid line given by Eq.~\eqref{eq:alpha_numerical}.

For $\varphi\sim D$, the ratio $\alpha$ scales as $\sim\sqrt{D}$, thus recovering the parametrically large gap in the non-sparse case.  For constant $\varphi$, the location of the bulk relative to its size is constant and independent of $D$, i.e, if measured relative to the size of the bulk, the bulk does not move away from the imaginary axis with increasing $D$.  We have thus quantified how sparsity cures one of the less physical aspects of the non-sparse random model of Markov generators.

\section{Spectral gap}\label{sec:gap}

\iffalse
\begin{figure}
	\begin{center}
		\includegraphics[width=\columnwidth]{gap.png}
		\caption{\label{fig:gap} Average spectral gap $\langle\gamma_*\rangle$ for edge weights distributed according to $\chi_2^2$ (top) and a uniform distribution on $[0,1]$ (bottom). Solid lines in the log-log plots are analytical predictions from Eq.~\eqref{eq:gap_mean_integral} in \textbf{(a)} and Eq.~\eqref{eq:gap_evt_uniform} in \textbf{(d)}. Black dashed lines in the heatmaps denote contours of constant gap. White circles in the heatmap in \textbf{(e)} are given by Eq.~\eqref{eq:const_gap_D}.}
	\end{center}
\end{figure}
\fi

\begin{figure*}
	\begin{center}
		\includegraphics[width=\textwidth]{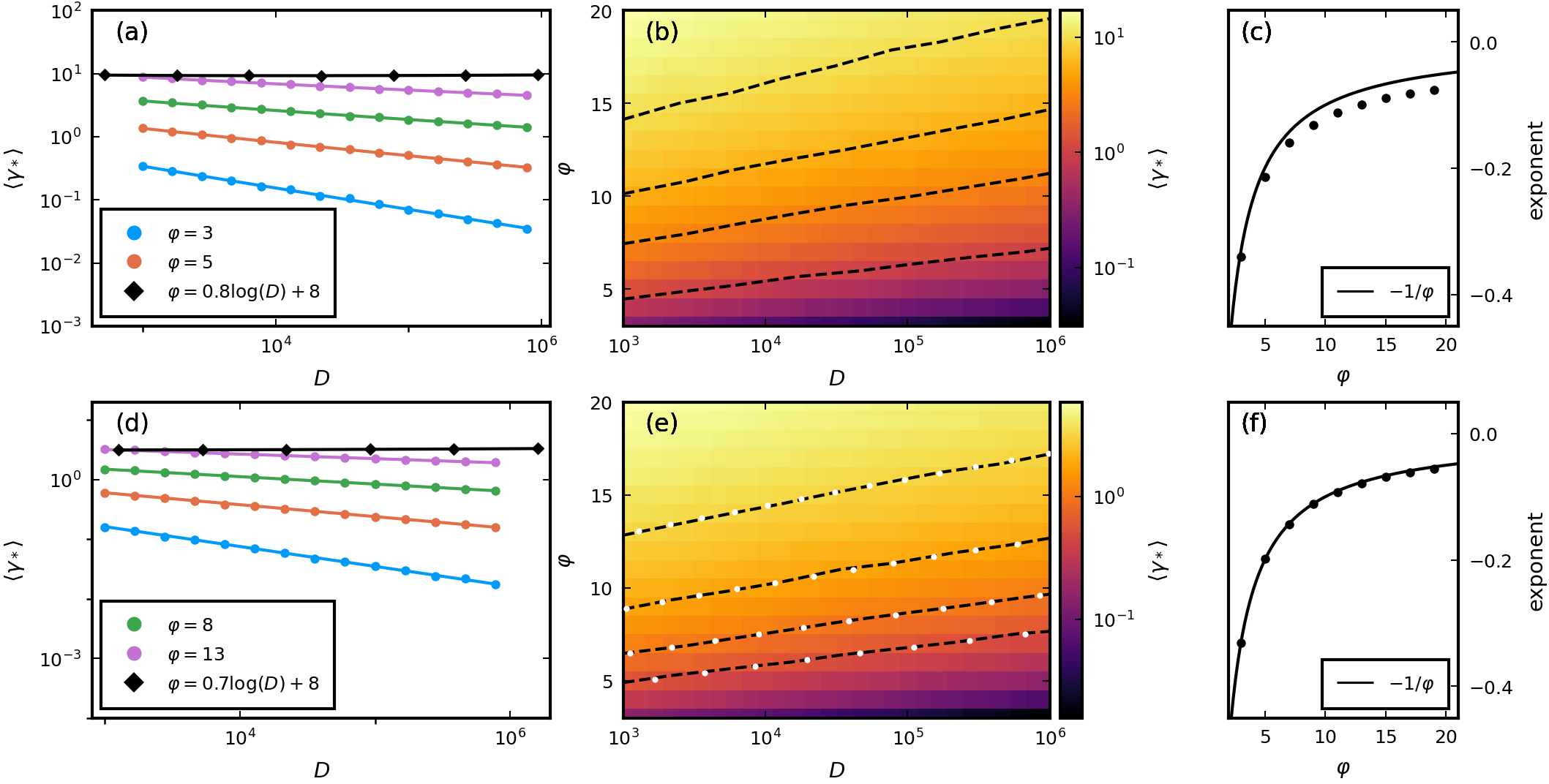}
		\caption{\label{fig:gap} The average spectral gap $\langle\gamma_*\rangle$ with $\chi_2^2$ (top) and standard uniform (bottom) weight distributions. Solid lines in the log-log plots are analytical predictions from Eq.~\eqref{eq:gap_mean_integral} in \textbf{(a)} and Eq.~\eqref{eq:gap_evt_uniform} in \textbf{(d)}. Black dashed lines in the heatmaps denote contours of constant gap. White circles in the heatmap in \textbf{(e)} are given by Eq.~\eqref{eq:const_gap_D}.}
	\end{center}
\end{figure*}

In this and the following section, we will consider the spectral edges, namely, the locations of the eigenvalues nearest and farthest from the imaginary axis.    
In this section we will investigate the spectral gap $\gamma_*$ of $\mathcal{K}$, 
\begin{equation}\label{eq:gap_def}
	\gamma_* = \min \{| \Re \lambda_i |: \Re \lambda_i < 0 \}.
\end{equation}
The spectral gap $\gamma_*$ is asymptotically, approximately bounded by the right extent of the bulk $|\mu(\lambda)| - \sigma(\lambda)$, which depends on $\varphi$ as $\sim \varphi - \sqrt{\varphi} \sim \varphi$. So for constant $\varphi$, the spectral gap is bounded from above, while for $\varphi$ increasing with $D$ the spectral gap can increase with $D$. 

Here the edge weights are distributed according to the $\chi_2^2$ and the standard uniform distributions. We first demonstrate that, for  $\varphi = \mathrm{const}$, the average spectral gap $\langle \gamma_*\rangle$ decreases as $D^{-1/\varphi}$, while $\langle \gamma_* \rangle$ is constant if $\varphi$ increases logarithmically with $D$. We then show that the spectral gap is well approximated by the smallest (in magnitude) diagonal term of $\mathcal{J}$ ($\mathcal{K}$) and use the theory of extreme values to underpin the numerical observations. The results are then generalized to weight distributions with power-law left tails in that for constant $\varphi$ the average spectral gap decreases as a power-law in $D$ and the crossover from decreasing to increasing $\langle \gamma_* \rangle$ happens when $\varphi\sim \log D$. 

\subsection{Numerical results}

In Figure~\ref{fig:gap} we show the average spectral gap $\langle \gamma_* \rangle$ for edge weights distributed as $\chi_2^2$ (a-c) and according to the standard uniform distribution (d-f). For every combination of $\varphi$ and $D$, the average of the spectral gap is estimated with $100$ samples. In Figure~\ref{fig:gap} (a) and (d) we show $\langle \gamma_* \rangle$ as a function of $D$ for different dependencies of $\varphi$ on $D$. The average spectral gaps for constant $\varphi=3,5,8,13$ (presented with colored circles) clearly follow a power-law scaling with $D$.

In Figure~\ref{fig:gap} (b) and (e) we show the average spectral gap $\langle \gamma_* \rangle$ as a function of $\varphi$  and $D$. The black dashed lines are contour lines of constant $\langle \gamma_* \rangle$. They are near straight lines, showing that for a logarithmic increase of $\varphi$ in $D$ the spectral gap is constant. 

We show the average spectral gap $\langle \gamma_*\rangle$ as a function of $D$ for $\varphi=\frac{4}{5}\log D + 8$ in Figure~\ref{fig:gap} (a) and $\varphi=\frac{7}{10}\log D + 8$ in (d) as black diamonds. These dependencies of $\varphi$ on $D$ agree well with the top dashed contour lines in (b) and (e), respectively. The average spectral gap of $\varphi$ depending logarithmically on $D$ is constant in Figure~\ref{fig:gap} (a) and (d).

\subsection{Gap $\approx$ minimum of $\mathcal{J}$}\label{sec:gap_gap_equals_J}
\begin{figure}
	\begin{center}
		\includegraphics[width=\columnwidth]{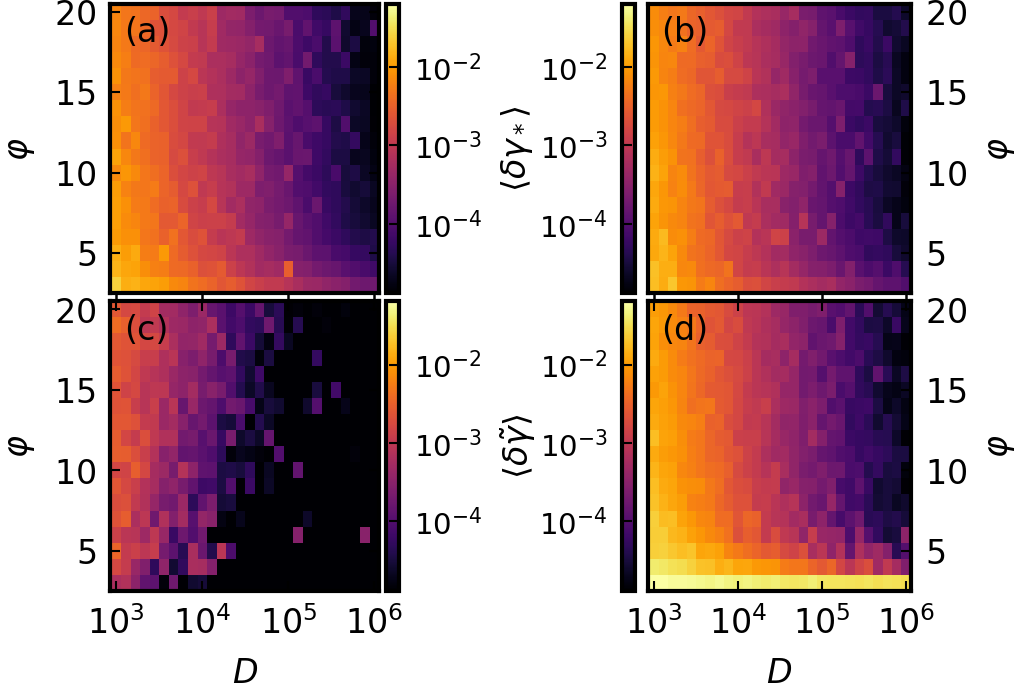}
		\caption{\label{fig:relative_error} The average relative error between the spectral gap and the minimal value of $\mathcal{J}$ in the top row (a) and (b) and between the horizontal extent and the maximum of $\mathcal{J}$ in the bottom row (c) and (d). 
  The weight distribution is $\chi_2^2$ on the left and the standard uniform distribution on the right. Averages are over 100 samples. See Eq.~\eqref{eq:gap_rel_err} and Eq.~\eqref{eq:extent_rel_err} for the definition of the relative errors $\delta \gamma_*$ and $\delta\tilde{\gamma}$, respectively. }
	\end{center}
\end{figure}

Let us assume for a moment that the generator matrix $\mathcal{K}$ is hermitian with eigenvalues $\lambda_D \le \dots \le \lambda_2 < \lambda_1 =0$. Then $\one = (1,\dots,1)^t$ is the eigenvector with eigenvalue 0 and all other eigenvectors are orthogonal to it. By the Courant-Fischer theorem \cite{Horn_Johnson_2012MatrixBook}
\begin{equation}\label{eq:ritz_gap}
	\gamma_* = - \lambda_2 = \min_{|v|=1, v \perp \one} v^t (-\mathcal{K}) v,
\end{equation}
where the minimum runs over all vectors $v\in\R^D$, which have Euclidean norm $|v|=1$ and are perpendicular to $\one$. Choosing $1\le l \le D$ arbitrary and $v$ as (see Appendix~\ref{sec:appendix_gap} for more details)
\begin{equation}
	v_i = \begin{cases}
		\sqrt{ 1 - \frac{1}{D}} &i=l \\
		-\frac{1}{\sqrt{D(D-1)}} &i\neq l,
	\end{cases}
\end{equation} 
a simple calculation shows that (at least for $\varphi\ll D$)
\begin{equation}\label{eq:gap_ineq_min}
	\gamma_* \le \min_{1\le l \le D} J_{ll} + O(D^{-1}).
\end{equation}
 Similarly, by using the Courant-Fisher theorem,  for the eigenvalue with largest magnitude $\lambda_D$ we find
\begin{equation}
	- \lambda_D = \max_{|v|=1} v^t (-\mathcal{K}) v,
\end{equation}
and with $v$ as the $l$-th vector of the standard basis of $\R^D$
\begin{equation}\label{eq:ineq_max}
	-\lambda_D \ge \max_{1\le l\le D} J_{ll}.
\end{equation}

Under some mild conditions on random weights $K_{ij}$, a result from Ref.~\cite{Bandeira_FoundCompMath2018} shows that the inequality Eq.~\eqref{eq:ineq_max} becomes an equality in the large $D$ limit with probability approaching 1. Motivated by this observation and the bound from Eq.~\eqref{eq:gap_ineq_min}, we expect a similar asymptotic tightness for Eq.~\eqref{eq:gap_ineq_min}. However, it is an open question whether the result from Ref.~\cite{Bandeira_FoundCompMath2018} applies to the bound of the spectral gap,  Eq.~\eqref{eq:gap_ineq_min}. Further, the proof presented in Ref.~\cite{Bandeira_FoundCompMath2018}
makes use of the Central Limit Theorem for the diagonal elements $J_{ll}$ of $\mathcal{J}$, and so the corresponding result does not apply to the case of constant or logarithmically  increasing (with  $D$) sparsity parameter $\varphi$.

Nevertheless, the above arguments allow us to conjecture that in the limit of large $D$ and $\varphi \ll D$ the following
\begin{equation}\label{eq:gap_equals_J}
	\gamma_* \approx \min_{1\le l \le D} J_{ll},
\end{equation}
holds for general, non-hermitian random generator matrices $\mathcal{K}$, with iid and non-exotic weight  distributions.  We support our conjecture with numerical data presented  in Figures~\ref{fig:relative_error} (a) and (b). We quantify the approximation in Eq.~\eqref{eq:gap_equals_J} by the relative error between the spectral gap $\gamma_*$ and the minimum $\min_{1\le l \le D} J_{ll}$ of the diagonal of $\mathcal{J}$,
\begin{equation}\label{eq:gap_rel_err}
	\delta \gamma_* = \frac{|\gamma_* - \min_{1\le l \le D} J_{ll}|}{ \gamma_*}.
\end{equation}
Figure~\ref{fig:relative_error} shows $\langle \delta \gamma_*\rangle$ as a function of $\varphi$ and $D$ for the $\chi_2^2$ distribution  and the standard uniform distribution. The average relative error is at least two orders of magnitude smaller than the average spectral gap shown in Figure~\ref{fig:gap} (b) and (e). For increasing $D$, the approximation in Eq.~\eqref{eq:gap_equals_J} improves. Thus, the approximation in Eq.~\eqref{eq:gap_equals_J}, works well in the case $\varphi\ll D$.
%and will be negligible in the large $D$ limit.

\subsection{Extreme value theory}\label{sec:gap_evt}
The distribution of the right-hand side of Eq.~\eqref{eq:gap_equals_J} can be tackled with the theory of extreme values. As all non-zero entries of $\mathcal{M}$ (edge weights) are identically and independently distributed random variables, so are the diagonal entries of $\mathcal{J}$. Let the cumulative distribution function (cdf) of the diagonal entries $J_{ll}$ of $\mathcal{J}$ be denoted by $F$ and its probability density function by $f(x) = \frac{d}{dx}F(x)$. If the edge weights are distributed according to a $\chi^2$ distribution (or any gamma distribution) the cdf $F$ of $J_{ll}$ is a gamma distribution function. If the edge weights are uniformly distributed, $F$ is an Irwin-Hall distribution function, see Table \ref{tab:C_beta}. The expected value of $\min_{1\le l \le D} J_{ll}$ is given in terms of $F$ (and $f$) by
\begin{equation}\label{eq:mean_integral_min}
	\left\langle \min_{1\le l \le D} J_{ll} \right\rangle 
	= D \int dx x f(x) (1-F(x))^{D-1}.
\end{equation}
Eq.~\eqref{eq:gap_equals_J} and Eq.~\eqref{eq:mean_integral_min} imply that
\begin{equation}\label{eq:gap_mean_integral}
	\langle \gamma_* \rangle \approx D \int dx x f(x) (1-F(x))^{D-1}.
\end{equation}

We demonstrate the validity of Eq.~\eqref{eq:gap_mean_integral} with Figure~\ref{fig:gap} (a), where the solid lines, given by Eq.~\eqref{eq:gap_mean_integral}, perfectly match  numerically sampled  average spectral gap $\langle \gamma_* \rangle$. In the next section, we will use the theory of extreme values to handle the integral in Eq.~\eqref{eq:gap_mean_integral}.

\subsubsection{Power-law tail distributions}
Let us consider first  the case $\varphi = \mathrm{const}$ and increasing $D$. By the Fisher-Tippet-Gnedenko (or 'extreme value') theorem~\cite{Embrechts_Kluppelberg_Mikosch_1997ExtremeEvents}, $\min_{1\le l \le D} J_{ll}$ converges in law, under some mild assumptions on the distribution of $J_{ll}$ and properly renormalization, to the Weibull distribution. The Weibull cumulative distribution function is given by $\Psi_\beta(x) = e^{-x^\beta}$, where $\beta>0$ and the support is on the positive real line. 

For distributions of $J_{ll}$ with power-law left tail, the renormalization of $\min_{1\le l \le D} J_{ll}$ for convergence to the Weibull distribution is well known, see e.g. Theorem 3.3.2, page 137 in Ref.~\cite{Embrechts_Kluppelberg_Mikosch_1997ExtremeEvents}. We use a version modified to our case. Let a positive random variable $X$ have cdf $F$ with $\beta$-power left tail, i.e. 
\begin{equation}
	F(x) = Cx^\beta \quad \text{for } 0\le x\le C^{1/\beta},
\end{equation}
where $C>0$ is a constant. Further, let $m_D = \min_{1\le l \le D} X_l$, where the $X_l$ are iid copies of $X$. Then
\begin{equation}\label{eq:Weibull_conv}
	(DC)^{1/\beta} m_D \to \Psi_\beta \quad \text{in law}.
\end{equation}
The Irwin-Hall distribution has a left power-law tail given by $F(x) = \frac{x^\varphi}{\varphi!}$ for $0\le x \le 1$. The  constants for the Irwin-Hall distribution are listed in table \ref{tab:C_beta}.

\begin{table}
\begin{center}
\begin{tabular}{ |c|c|c| }
	\hline
	off-diag. $\mathcal{K}=M_{ij}$	& $\chi_k^2$ 								& uniform \\ 
	diag. $\mathcal{K}=J_{ll}$		& gamma$\left(\frac{k\varphi}{2},2\right)$ 	& Irwin-Hall \\ 
	\hline
	C 							& $\frac{2^\varphi}{\varphi!}$* 	& $\frac{1}{\varphi!}$ \\ 
	$\beta$ 					& $\frac{k}{2}\varphi$*				& $\varphi$	\\ 
	\hline
\end{tabular}
\end{center}
\caption{\label{tab:C_beta} The distributions of the off-diagonal elements $M_{ij}$ of $\mathcal{K}$ (edge weights) and the corresponding distributions of the diagonal elements $J_{ll}$ of $\mathcal{K}$ and the corresponding constants $C$ and $\beta$ for the convergence of $J_{ll}$ to the Weibull distribution $\Psi_\beta$ in Eq.~\eqref{eq:Weibull_conv}. (*) constants obtained by a power-law approximation of the left tail of the gamma distribution.}
\end{table}

We assume that the convergence in Eq.~\eqref{eq:Weibull_conv} is not only in distribution but that the renormalized moments of $m_D$ converge as well. If the convergence of the moments is sufficiently fast, then Eq.~\eqref{eq:Weibull_conv} together with Eq.~\eqref{eq:gap_equals_J} imply
\begin{equation}\label{eq:gap_evt_uniform}
	\langle \gamma_* \rangle 
	\approx \langle m_D \rangle 	
	\approx \Gamma\left(1+\frac{1}{\varphi}\right) (\varphi!)^{1/\varphi} D^{-1/\varphi}
\end{equation}
when the weight distribution (distribution of non-zero off-diagonal elements of $\mathcal{K}$) is  such that the diagonal of $\mathcal{J}$ has a power-law left tail and the coefficients $C$ and $\beta$ are given by $C=1/\varphi!$ and $\beta=\varphi$. 

Finally, we consider the case that the weight distribution is uniform. We observe that the approximation in Eq.~\eqref{eq:gap_evt_uniform} works very well in this case. The solid lines in Figure~\ref{fig:gap} (d) are given by the right-hand side of Eq.~\eqref{eq:gap_evt_uniform} and they match the numerically calculated average spectral gap.

Eq.~\eqref{eq:gap_evt_uniform} implies that, for constant $\varphi = \mathrm{const}$ and increasing $D$, the average spectral gap decreases as 
%a power of $D$ with exponent $-1/\varphi$,
\begin{equation}\label{eq:gap_phi_const_uniform}
	\langle \gamma_* \rangle \sim D^{-1/\varphi}.
\end{equation}
In Figure~\ref{fig:gap} (f) we show that the numerically retrieved power-law exponents of the average spectral gap, Figure~\ref{fig:gap} (d), match the scaling in Eq.~\eqref{eq:gap_phi_const_uniform}.

We find that the large deviation result is not only valid for constant $\varphi$ and increasing $D$ but also for $\varphi$ increasing logarithmically with $D$; see Figure~\ref{fig:gap} (d). This allows us to estimate the crossover from decreasing to increasing spectral gap. Let $c$ denote a constant and let $\langle \gamma_*\rangle = c$. Then by Eq.~\eqref{eq:gap_evt_uniform}
\begin{equation}\label{eq:const_gap_D}
	D \approx \left[ \frac{\Gamma\left( 1 + \frac{1}{\varphi}\right)}{c} \right]^{\varphi} \varphi!.
\end{equation}
In Figure~\ref{fig:gap} (e) the contour lines of constant average spectral gap $c$ perfectly line up with the functional dependence of $D$ on $\varphi$ through Eq.~\eqref{eq:const_gap_D} shown as white dots.

To find $\varphi$ as a function of $D$ such that the average spectral gap is constant, we assume that $\varphi$ is reasonably large and approximate $\Gamma\left(1+\frac{1}{\varphi}\right)\approx 1$ and by Stirling's formula $(\varphi!)^{1/\varphi}\approx \frac{\varphi}{e}$. Denoting $y = \log \frac{\varphi}{ce}$ and rearranging Eq.~\eqref{eq:const_gap_D} gives us
\begin{equation}
	\frac{\log D}{ce}  \approx ye^y,
\end{equation}
which can be inverted by the Lambert $W$ function. Resubstituting $\varphi= cee^y$ we arrive at
\begin{equation}
	\varphi 
	\approx ce \cdot e^{W\left(\frac{\log D}{ce}\right)},
\end{equation}
which for $\log D \ge ce^2$ behaves as \cite{Hoorfar_Hassini_JourInequMath2008}
\begin{equation}\label{eq:const_gap_phi}
	\varphi \approx \frac{\log D}{\left(\log \log D - \log c - 1\right)^{1-\eta(D)}},
\end{equation}
where $\eta(D) \to 0$ slowly, as $\eta(D) \sim (\log\log D)^{-1}$. So in the limit $1\ll \varphi \ll D$ the crossover from decreasing to increasing spectral gap happens at $\varphi \sim \log D$ with corrections of the order $\log\log D$. This confirms our numerical observations that the average spectral gap $\langle \gamma_* \rangle$ appears to be constant for $\varphi\sim\log D$ in the range of matrix sizes $D$ we considered.

\subsubsection{Approximate power-law distributions}
If the weight distribution is a $\chi^2$ or exponential distribution, the diagonal elements of $\mathcal{J}$ are distributed according to Gamma distribution, see table \ref{tab:C_beta}. The left tail of the Gamma distribution only follows approximately a power-law. Approximating the left tail by a Taylor expansion, we obtain constants $C$ and $\beta$ presented in Table \ref{tab:C_beta}. Especially, for the $\chi_2^2$ distribution, we presented so far in the main text the power-law approximation of the gamma distribution and the large deviation result in the previous subsection suggest that the average spectral gap $\langle \gamma_* \rangle$ decreases for constant $\varphi$ and increasing $D$ as a power in $D$ with exponent given $-1/\varphi$, see Eq.~\eqref{eq:gap_phi_const_uniform}. 

In Figure~\ref{fig:gap} (c) we present the numerically calculated exponents of the power-law decrease of $\langle \gamma_* \rangle$, for $\chi_2^2$ weight distribution, with $D$ and compare it to the prediction $-1/\varphi$. We find excellent agreement for small $\varphi \le 5$. For larger $\varphi$ the deviation between the numerical exponent and $-1/\varphi$ is visible but the agreement is still good.

A quantitative comparison between the numerically calculated spectral gap $\langle \gamma_* \rangle$ and the EVT prediction by a power-law approximation of the left tail of the gamma distribution resulted in poor agreement. As the expected minimum value of the diagonal of $\mathcal{K}$ perfectly agrees with $\langle \gamma_*\rangle$, we attribute the disagreement to the power-law approximation of the left tail and slow convergence of Eq.~\eqref{eq:Weibull_conv} for  diagonal elements of $\mathcal{J}$ distributed according to the gamma distribution.

\subsection{Summary}
We presented numerical and analytical arguments that, for the weight distributions considered, the average spectral gap decreases as a power-law for constant $\varphi$ and increasing $D$ with exponent given (approximately) by $-1/\varphi$. The crossover between decreasing and increasing spectral gap happens at $\varphi\sim\log D$, with $\log\log D$ corrections, for uniform weight distribution. For $\chi_2^2$ distributed edge weights the crossover was observed at $\varphi\sim\log D$. If $\varphi$ increases with $D$ faster than $\log D$ then the average spectral gap increases.

The presented results generalize. Let us assume that the spectral gap is well approximated by the smallest (in magnitude) diagonal of $\mathcal{J}$, at least in the regime of large $D$ and $\varphi\ll D$. Then, after appropriate renormalization, the distribution of the spectral gap is given by the limiting extreme value distribution of the diagonal elements of $\mathcal{J}$. Thus the classification of functional dependencies of the spectral gap on $\varphi$ and $D$ with respect to weight distributions reduces to the classification of limiting extreme value distributions and renormalizations. Extensive research has been conducted on the latter and the renormalizations of a lot of common distributions are well known \cite{deHaan_Ferreira_2006EVT, Embrechts_Kluppelberg_Mikosch_1997ExtremeEvents}. Thus the presented approach allows the calculation of the distribution of the spectral gap for broad classes of weight distributions.

\section{Horizontal extent (largest absolute real part)}\label{sec:extent}

\begin{figure}
	\begin{center}
		\includegraphics[width=\columnwidth]{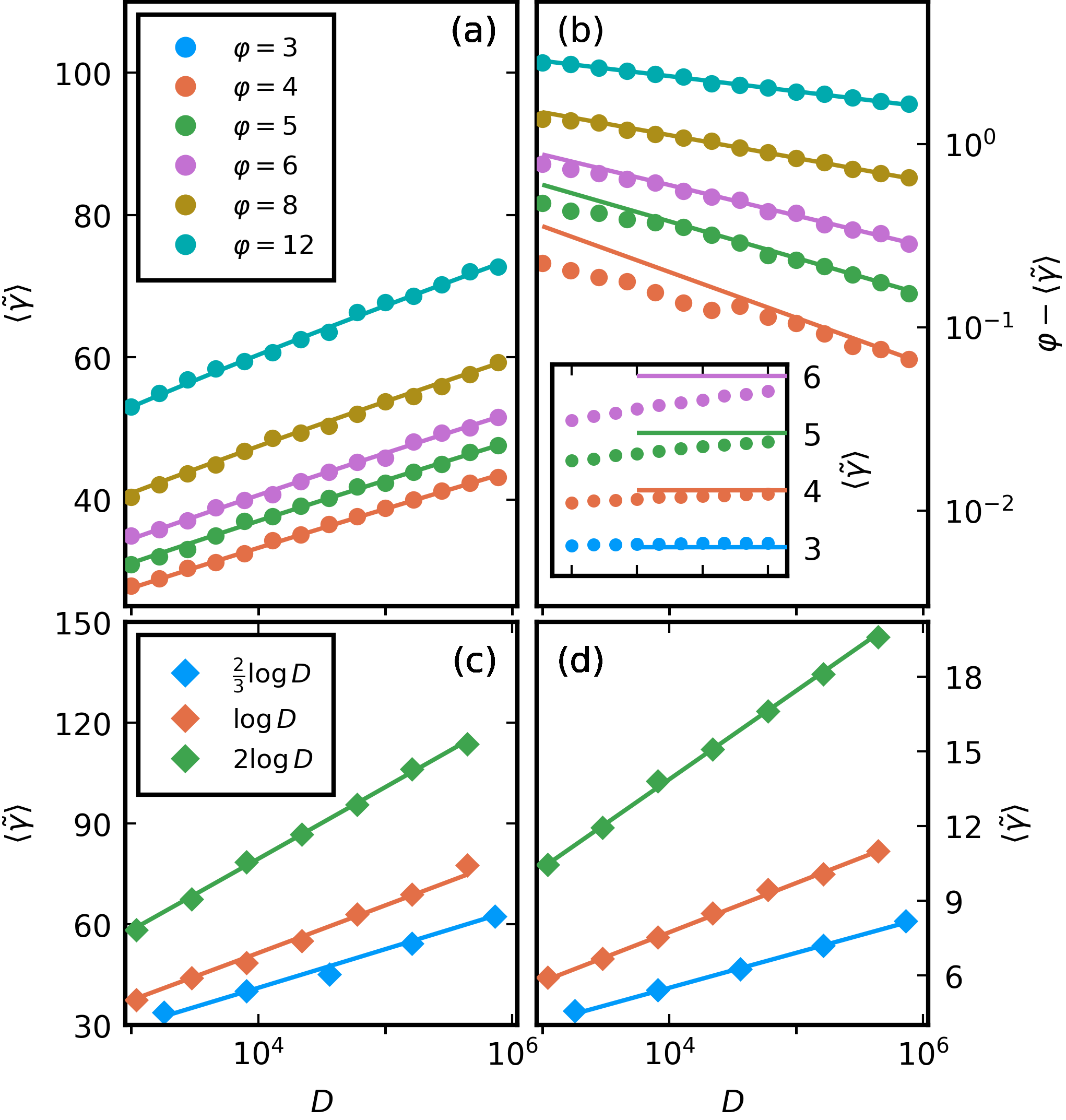}
		\caption{\label{fig:extent} Average horizontal extent $\langle \tilde{\gamma} \rangle$ with $\chi_2^2$ (left) and standard uniform (right) weight distributions. $\varphi$ is constant (top) and $\varphi\sim \log D$ (bottom). Solid lines are given by Eq.~\eqref{eq:extent_mean_integral} (left) and Eq.~\eqref{eq:extent_evt_uniform} (right).}
	\end{center}
\end{figure}

In this section we investigate the horizontal extent $\tilde{\gamma}$ of the spectrum given by the eigenvalue with largest absolute real part
\begin{equation}
	\tilde{\gamma} = \max_{1\le i\le D} |\Re \lambda_i|.
\end{equation}
We focus on the averaged horizontal extent $\langle \tilde{\gamma} \rangle$. We show that for $\varphi\sim\log D$ the average horizontal extent increases logarithmically with $D$ for $\chi_2^2$ or uniformly distributed edge weights. For constant $\varphi$ and increasing $D$ the dependence of $\langle \tilde{\gamma} \rangle$ is qualitatively very different for the two distributions. For the $\chi_2^2$ distribution $\langle \tilde{\gamma} \rangle$ increases logarithmically, while for the uniform distribution, the average horizontal extent converges to $\varphi$ as a power-law in $D$. Ultimately, this is because the support of the uniform distribution is bounded, while the right tail of the $\chi_2^2$ distribution extends to infinity.

The structure of this section follows closely the one from Section~\ref{sec:gap}. We first present numerical results demonstrating the above statements. We then argue that the horizontal extent is given by the largest, in magnitude, diagonal element of $\mathcal{K}$ and invoke again EVT to analytically underpin the functional dependencies of $\langle \tilde{\gamma} \rangle$ on $\varphi$ and $D$.

\subsection{Numerical results}

In Figure~\ref{fig:extent} we show the average horizontal extent as a function of $D$ for constant $\varphi$ and $\varphi\sim\log D$ for edge weights distributed according to a $\chi_2^2$ (a,c) and the standard uniform distribution (b,d). In (a) the dependence of $\langle \tilde{\gamma} \rangle$ on $D$ for constant $\varphi$ shows a clear logarithmic increase with $D$ for $\chi_2^2$ distributed edge weights. In contrast, for the uniform distribution, the average horizontal extent increases with $\varphi$ as a power-law, see (b). The power-law behavior sets in for small $\varphi$ only for larger $D$. For $\varphi=4$ and $\varphi=5$ deviations from the straight lines in Figure~\ref{fig:extent} (b) are visible for $D<10^5$ and $D<10^4$, respectively. The average horizontal extent for constant $\varphi=2$ and $\varphi=3$ is not shown. We found that it does not converge to $\varphi$ in the range of matrix sizes $D$ we investigated. 

For $\varphi\sim\log D$ the dependence of $\langle \tilde{\gamma}\rangle$ on $D$ is logarithmic for both the $\chi_2^2$ and the uniform distribution, as shown in Figure~\ref{fig:extent} (c) and (d).

In the remainder of this section, we will present analytic arguments similar to Section~\ref{sec:gap}. We will explain the difference of the dependence of $\langle \tilde{\gamma} \rangle$ on $D$ for constant $\varphi$ between $\chi_2^2$ and uniform-like distributions. We show that $\llangle \tilde{\gamma} \rrangle \sim\log D$ for both distributions and $\varphi\sim \log D$.

\subsection{Extent $\approx$ maximum of $\mathcal{J}$}
By the Perron-Frobenius theorem the spectrum of $\mathcal{K}$ is confined to the ball centered around $\min_{i} K_{ii} < 0$ with radius $r=|\min_{i} K_{ii}|$. Thus $2 \max_{1\le l \le D} J_{ll} \ge |\Re \lambda|$ for all eigenvalues $\lambda$, so
\begin{equation}\label{eq:extent_bound_perron}
	\tilde{\gamma} \le 2 \max_{1\le l \le D} J_{ll}.
\end{equation}

For symmetric generator matrices $\mathcal{K}$ we showed in Section~\ref{sec:gap_gap_equals_J} that 
\begin{equation}
	\max_{1\le l \le D} J_{ll} \le \tilde{\gamma}
\end{equation}
and stated a result from \cite{Bandeira_FoundCompMath2018} that for symmetric random generator matrices under mild conditions on the weights $K_{ij}$, $\max_{1\le l \le D} J_{ll}$ concentrates around the largest eigenvalue in magnitude, $\tilde{\gamma}$. This together with the upper bound by the Perron-Frobenius theorem Eq.~\eqref{eq:extent_bound_perron} leads to our conjecture that the concentration of $\max_{1\le l \le D} J_{ll}$ around $\tilde{\gamma}$ in the symmetric case extends to the non-hermitian case as well
\begin{equation}\label{eq:extent_equals_J}
	\tilde{\gamma} \approx \max_{1\le l \le D} J_{ll}.
\end{equation}
A concentration result similar to the one in \cite{Bandeira_FoundCompMath2018} for non-hermitian random generator matrices $M$ has to the best of our knowledge not appeared in the literature.

To quantify the deviation in Eq.~\eqref{eq:extent_equals_J} we introduce the relative error of $\tilde{\gamma}$ and $\max_{1\le l \le D} J_{ll}$
\begin{equation}\label{eq:extent_rel_err}
	\delta \tilde{\gamma} = \frac{|\tilde{\gamma} - \max_{1\le l\le D} J_{ll}|}{\tilde{\gamma}}.
\end{equation}
In Figure~\ref{fig:relative_error} (c) and (d) we show the average relative error $\langle \delta \tilde{\gamma} \rangle$ as a function of $D$ and $\varphi$. If the edge weights are $\chi_2^2$ distributed then for $2\le \varphi \le 20$ and $10^3 \le D \le  10^5$ the average relative error is smaller than $\approx 10^{-3}$ and decreases with increasing $D$. Thus Eq.~\eqref{eq:extent_equals_J} is a good approximation for large $D$ and $\varphi \ll D$ and the error appears negligible in the limit of large $D$. For uniformly distributed edge weights the average relative error $\langle \delta\tilde{\gamma} \rangle$ is smaller than $10^{-1}$ for $2\le \varphi \le 20$ and $10^3 \le D \le 10^5$ and for $\varphi \ge 4$ decreases with $D$. For $2\le \varphi \le 3$, the error does not seem to decrease for increasing $D$. We conclude that Eq.~\eqref{eq:extent_equals_J} is an excellent approximation for large $D$ and $4\le \varphi \ll D$.

\subsection{Extreme value theory}
Recall that the diagonal elements of $\mathcal{J}$ are iid random variables. Similar to the minimum extreme value statistics, the expected value of $\max_{1\le l \le D} J_{ll}$ is
\begin{equation}\label{eq:extent_mean_integral}
	\left\langle \max_{1\le l \le D} J_{ll} \right\rangle 
	= D \int dx x f(x) F(x)^{D-1},
\end{equation}
where we denoted the cdf of the diagonal elements $J_{ll}$ of $\mathcal{J}$ by $F$ and the pdf by $f=\frac{d}{dx} F$. A numerical calculation of the integral in Eq.~\eqref{eq:extent_mean_integral} for $\chi_2^2$ distributed edge weights is shown in Figure~\ref{fig:extent} (a) and (c) and compared to the average horizontal extent $\langle\tilde{\gamma}\rangle$. The quantities agree excellently. 

The remainder of this section is devoted to employing the Fisher-Tippet Gnedenko or extreme value theorem to $\max_{1\le l \le D} J_{ll}$ and thus analytically calculate the integral in Eq.~\eqref{eq:extent_mean_integral}.
%, when the off-diagonal elements of $M$ are distributed according to a $\chi^2$ or uniform distribution.

\subsubsection{Gamma distribution}

\begin{table}
\begin{center}
\begin{tabular}{ |c|c| }
	\hline
	& gamma$(k, \theta)$ \\
	$c$		& $\frac{1}{\theta}$ \\
	$d(D)$	& $\theta(\log D + (k-1)\log\log D - \log \Gamma(k))$ \\
	\hline
	$\chi_n^2$ 			& gamma$(k=\frac{n}{2} \varphi, \theta=2)$ \\
%	expon.$(\lambda)$ 	& gamma$(k=\varphi, \theta=\lambda)$ \\
	\hline
\end{tabular}
\end{center}
\caption{\label{tab:c_D} (top) The normalizing parameters $c$ and $d(D)$ for $\max_{1\le l \le D} J_{ll}$ to converge to the Gumbel distribution, where $J_{ll}$ is gamma distributed with shape and rate parameter $k$ and $\theta$, respectively, see Eq.~\eqref{eq:Gumbel_convergence}. (bottom) The relation between the $\chi^2$ and the gamma distribution.}
\end{table}

Recall that if the edge weights are distributed according to a $\chi^2$ distribution then the diagonal elements of $\mathcal{J}$ are gamma distributed. The maximum of $D$ gamma distributed iid random variables $X_l$ converges in law to a standard Gumbel distribution $\operatorname{Gum}$ \cite{Bandeira_FoundCompMath2018},
\begin{equation}\label{eq:Gumbel_convergence}
	c\left[\max_{1\le l\le D} X_l - d(D)\right] \to \operatorname{Gum} \quad \text{in law},
\end{equation}
with parameters $c$ and $d(D)$ given in table \ref{tab:c_D} for the gamma and $\chi^2$ distributions. The cdf of the Gumbel distribution is $x\to e^{-e^{-x}}$ with mean $\gamma$, where $\gamma$ denotes the Euler-Mascheroni constant, not to be confused with the horizontal extent $\tilde{\gamma}$. The assumption that the first moment converges and the convergence is fast enough together with Eq.~\eqref{eq:extent_equals_J} yields
\begin{equation}\label{eq:extent_evt_gamma}
	\langle \tilde{\gamma} \rangle 
	\approx \left\langle \max_{1\le l \le D} J_{ll} \right\rangle 
	\approx \frac{\gamma}{c} +d(D).
\end{equation}

For constant $\varphi$ and increasing $D$ the dominant contribution of $d(D)$ is $2\log D$ for the $\chi^2$ distribution. Thus the increase is expected to be logarithmic. This is qualitatively consistent with numerical calculations of the average horizontal extent of random generator matrices $\mathcal{K}$ with $\chi_2^2$ distributed edge weights and constant $\varphi$ shown in Figure~\ref{fig:extent} (a). There $\tilde{\gamma}$ increases logarithmically with $D$. Quantitatively, the deviation between the average horizontal extent and the right-hand side of Eq.~\eqref{eq:extent_evt_gamma} is not small. The deviation decreases for increasing $D$ (not shown). We attribute the slow convergence to a sub-optimal choice of parameters $c$ and $d(D)$, as the right-hand side of Eq.~\eqref{eq:extent_mean_integral} agrees perfectly with the numerically calculated $\langle \tilde{\gamma} \rangle$.

Let us assume that Eq.~\eqref{eq:extent_evt_gamma} is valid for $\varphi$ increasing logarithmically. Note that for the $\chi^2$ distribution, the rate parameter of the corresponding gamma distribution is linear in $\varphi$. Thus for large enough $\varphi$ by Stirling's formula, the dominant term in Eq.~\eqref{eq:extent_evt_gamma} is logarithmic in $D$. Hence the average horizontal extent $\langle\tilde{\gamma} \rangle$ should increase logarithmically for $\varphi\sim\log D$. This is again qualitatively confirmed by numerical results shown in Figure~\ref{fig:extent} (c), where $\langle \tilde{\gamma} \rangle$ as a function of $D$ for $\varphi\sim\log D$ increases logarithmically with $D$.

\subsubsection{Power-law tail distributions with bounded support}
If the distribution of the diagonal of $\mathcal{J}$ has bounded support and the right tail decreases as a power-law, then we can reuse the extreme value result from Section~\ref{sec:gap_evt}. For a random variable $X$ with right support endpoint $x_0$ and cdf $F$ with power-law right tail, i.e.
\begin{equation}
	F(x) = C (x_0-x)^\beta\quad \text{for } x_0-C^{1/\beta} \le x \le x_0,
\end{equation}
then $m_D = \max_{1\le i\le D} X_l$, where $X_l$ are iid copies of $X$, converges, properly renormalized, in law to a Weibull distribution
\begin{equation}
	(DC)^{1/\beta} (x_0 - m_D) \to \Psi_\beta\quad \text{in law}.
\end{equation}

Again, assuming that the first moment converges as well and the convergence is fast enough we get for edge weights distributed according to the standard uniform distribution,
\begin{equation}\label{eq:extent_evt_uniform}
	\langle \tilde{\gamma} \rangle 
	\approx \langle \max_{1\le l \le D} J_{ll} \rangle
	\approx \varphi - \Gamma\left(1+\frac{1}{\varphi}\right) (\varphi!)^{1/\varphi} D^{-1/\varphi}.
\end{equation}
We find excellent numerical agreement of the right-hand side of Eq.~\eqref{eq:extent_evt_uniform} with the average horizontal extent $\langle \tilde{\gamma} \rangle$ for $\varphi \ge 4$. In Figure~\ref{fig:extent} (b) we show $\langle \tilde{\gamma} \rangle$ as a function of $D$ for fixed $\varphi$. The solid lines denote the right-hand side of Eq.~\eqref{eq:extent_evt_uniform}. They agree perfectly for large enough $D$ and $\varphi \ge 4$. For $4\le \varphi\lessapprox 6$ and small $D$ the agreement is still reasonable but deviations are clearly visible. Thus for fixed $\varphi\ge 4$ and increasing $D$, $\langle \tilde{\gamma}\rangle$ converges to $\varphi$ as a power in $D$ with exponent $-1/\varphi$,
\begin{equation}
	\varphi - \langle \tilde{\gamma} \rangle\sim D^{-1/\varphi}.
\end{equation}

Numerically we find that Eq.~\eqref{eq:extent_evt_uniform} is valid for $\varphi$ increasing with $D$ logarithmically, see Figure~\ref{fig:extent} (d). There we show the average horizontal extent as a function of $D$ for $\varphi=\log D$. It increases logarithmically with $D$. The logarithmic increase can be justified analytically by extending Eq.~\eqref{eq:extent_evt_uniform} beyond constant $\varphi$. In the limit of large enough $\varphi$ we approximate $\Gamma(1+1/\varphi) \approx 1$ and by Stirling's formula $(\varphi!)^{1/\varphi}\approx \varphi/e$ and get
\begin{equation}
	\langle \tilde{\gamma} \rangle \approx \varphi (1-D^{-1/\varphi}) \sim \varphi.
\end{equation}
Thus in the limit of large $\varphi$ the average horizontal extent increases as $\sim \varphi\sim \log D$.

\subsection{Summary}
We showed numerically and analytically that the horizontal extent increases logarithmically for $\chi_2^2$ and uniformly distributed edge weights if $\varphi\sim\log D$. For constant $\varphi\ge4$ and uniformly distributed edge weights the horizontal extent increases to $\varphi$ as $\sim\varphi-D^{-1/\varphi}$, while $\langle\tilde{\gamma}\rangle$ increases logarithmically for constant $\varphi$ and $\chi_2^2$ distributed edge weights.

The difference of the dependence of the average horizontal extent on $\varphi$ between the $\chi_2^2$ and uniform distribution goes back to the difference of the right tails. When edge weights are uniformly distributed the diagonal has bounded support and a power-law right tail, while it has unbounded support and an exponentially decaying right tail for $\chi_2^2$ distributed edge weights.

Similar to the spectral gap the limiting distribution of the horizontal extent is given by the limiting extreme value distribution of the diagonal elements of $\mathcal{K}$, under the assumption that the largest (in magnitude) diagonal of $\mathcal{K}$ is approximating $\tilde{\gamma}$ well enough. Thus the classification of the horizontal extent with respect to weight distributions reduces to the classifications of convergence in extreme value theory.

\section{Complex Spacing ratios}\label{sec:ratios}

\begin{figure}
\begin{center}
	\includegraphics[width=\columnwidth]{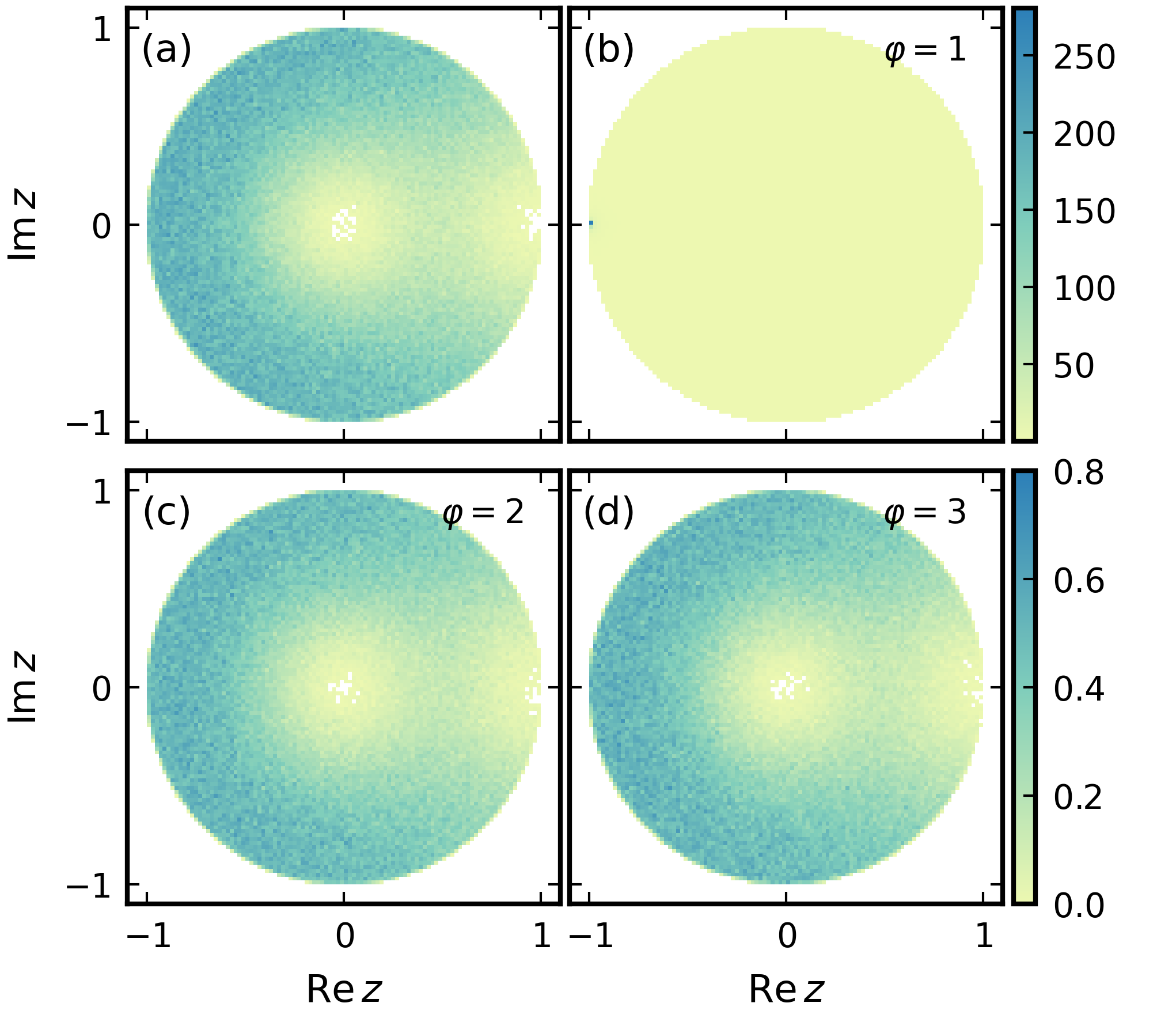}
	\caption{\label{fig:spacing_ratios} Density of complex spacing ratios for \textbf{(a)} real Ginibre ensemble and \textbf{(b)-(d)} sparse Kolmogorov operators with $\varphi=1,2,3$. The number of states $D=10^4$ and densities are obtained from $10^2$ samples. Edge weights are distributed according to the $\chi_2^2$ distribution. The color range is from 0 to 0.8 in (a), (c), and (d) and from 0 to 260 in (b).}	
\end{center}
\end{figure}

So far we considered the marginal distribution of eigenvalues of sparse random generator matrices. But correlations between the eigenvalues are also of interest. Correlations between eigenvalues of real spectra are often quantified with the distribution of consecutive level spacings or their ratios. The latter  avoids the need to unfold the corresponding spectrum \cite{Vadim_Huse_PRB2007, Atas_et_al_PRL2013ratios} and has been generalized to complex eigenvalues in the recent work~\cite{Sa_Ribeiro_Prosen_PRX2020}. The complex spacing ratio (CSR) of  eigenvalue $\lambda$ of  matrix $\mathcal{K}$ is defined as
\begin{equation}\label{eq:spacing_ratio}
	z = \frac{\lambda^{NN} - \lambda}{\lambda^{NNN} -\lambda},
\end{equation}
where $\lambda^{NN}$ and $\lambda^{NNN}$ denote the closest, by the Euclidean distance, and second closest eigenvalue (of $\mathcal{K}$) to $\lambda$, respectively. By definition, the density of CSRs is supported on the unit disk on the complex plane. 

If eigenvalues $\lambda$ are uncorrelated, the CSR density is uniform. Eigenvalues of generic random matrix ensembles are typically correlated and feature mutual repulsion. This leads to vanishing CSR density at $z=0$ and $z=1$. According to \cite{Ueda_PRR2020universalityclass}, complex level spacings categorize random matrix ensembles in three universality classes. Generic random matrices fall into one of these classes according to their symmetries. The random generators considered in this paper have real entries so they obey the same symmetry as real Ginibre matrices (GinOE).

In Figure~\ref{fig:spacing_ratios} we show the CSR densities of (a) GinOE members with Gaussian entries (b-d) and sparse random generators with $\chi_2^2$ distributed edge weights and $\varphi=1,2,3$. The densities are estimated from $100$ samples for $D=10^4$. We also checked that the obtained densities are independent of the weight distribution. As suggested in Ref.~\cite{Sa_Ribeiro_Prosen_PRX2020}, we avoid eigenvalues close to the real line (by excluding all eigenvalues from the strip $\Im \lambda < 10^{-14}$) when sampling CSR densities. 

The CSR density of GinOE matrices shown on Figure~\ref{fig:spacing_ratios} (a) exhibit  typical depletion at $z=0$ and $z=1$. In Ref.~\cite{Tarnowski_Denisov_etal_PRE2021}, it was shown that the CSR density obtained for dense random Kolmogorov operators  agrees well with the distribution shown in Figure~\ref{fig:spacing_ratios} (a). The CSR density of sparse generators with sparsity $\varphi\ge 2$ (c,d) agrees remarkably well with the GinOE case.  

The CSR density for $\varphi=1$ is anomalous, see Figure~\ref{fig:spacing_ratios} (b). It has an extremely high density around $z=-1$ while being nearly flat on the rest of the unit disk.
In this ultimate case, the operator can be 
presented as
\begin{equation}\label{eq:Hatano}
{\cal K} = {\cal V} \cdot({\cal P}-\oper),
\end{equation}
where ${\cal V}$ is  a diagonal matrix (with elements distributed according to, e.g., $\chi_2^2$) and ${\cal P}$ is a circulant permutation matrix corresponding to a cyclic unit shift. The spectrum of ${\cal P}-\oper$ lies on a circle of unit radius centered at $\lambda=-1$ and constitutes a set of equidistant roots of unity. This spectrum is slightly deformed and split into several loops by multiplication of ${\cal P}-\oper$ with ${\cal V}$. Away from $\lambda =-1$, ${\cal V}$ dominates which results in the appearance of a real 'tail'; see Fig. 8. 

In graph theory terms, such a sparse random graph fragments into a set  of disjoint elementary cycle graphs.  The independence of the spectra of these cycles leads to the flatness of the density away from $z=-1$, while the elementary  cycle structure of the connected components is responsible for the CSR peak at $z=-1$.

\begin{figure}[b]
\begin{center}
	\includegraphics[width=\columnwidth]{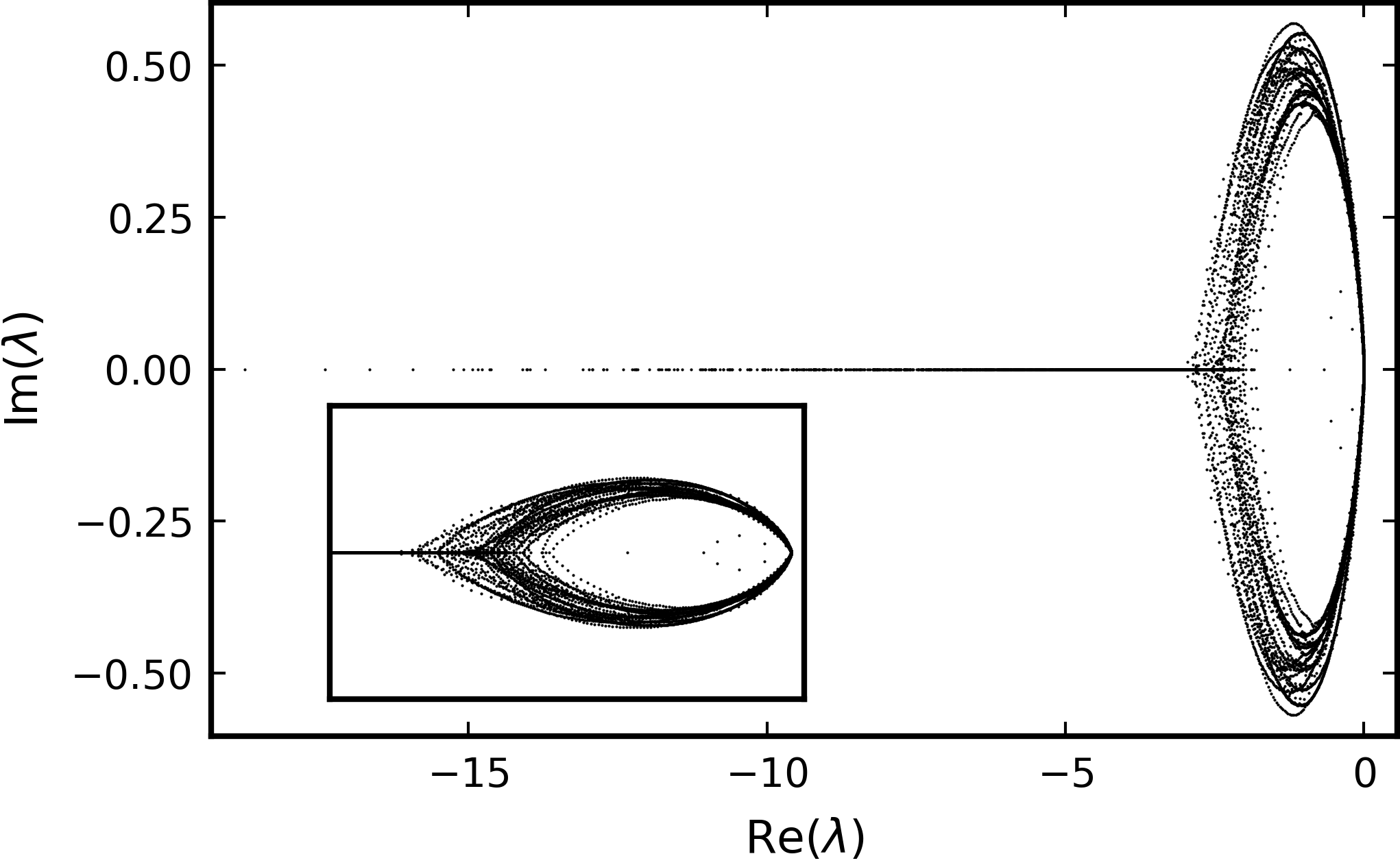}
	\caption{\label{fig:sloops} { Spectrum  of a random Kolmogorov operator with $\varphi=1$ and $\chi_2^2$ weight distribution. The matrix size is $D=10^3$. Inset: same data plotted with both axes having the same scale.}}	
\end{center}
\end{figure}

To quantify the distance between CSR densities, we use the average length $\langle r \rangle$ and the average cosine of the angle $-\langle \cos \theta \rangle$ of spacing ratios, where $\langle \dots\rangle$ again denotes the average over the random matrix ensemble~\cite{Sa_Ribeiro_Prosen_PRX2020}. 
%As analytical calculations of length and angle of GinOE spacing ratios have to the best of our knowledge not appeared in the literature, 
We numerically estimate $\langle r \rangle_{\ginoe}\approx0.7379$ and $- \langle \cos \theta \rangle_{\ginoe}\approx0.2347$ for 100 $10^4\times10^4$-matrices. These agree well with $\langle r\rangle$ and $-\langle\cos\theta\rangle$ for $\varphi=2$ and $\varphi=3$, as shown in Table \ref{tab:ratios}. We found similar results for $\varphi>3$ (not shown). In contrast, the corresponding quantities for  $\varphi=1$ deviate substantially  from $\langle r \rangle_{\ginoe}$ and $-\langle \cos \theta \rangle_{\ginoe}$, as also shown in Table \ref{tab:ratios}. We conclude that, for $\varphi\ge 2$, correlations between eigenvalues of sparse random Kolmogorov operators agree with correlations of eigenvalues  of GinOE matrices. 

\begin{table}
\begin{center}
	\begin{tabular}{ |c|c|c|c|c| }
		\hline
										& GinOE & $\varphi=1$ & $\varphi=2$& $\varphi=3$ \\
		\hline
		$-\langle \cos \theta \rangle$ 	& 0.7379 & 0.7871 & 0.7359 & 0.7372 \\
		$\langle r\rangle$ 				& 0.2347 & 0.3516 & 0.2225 & 0.2284 \\
		\hline
	\end{tabular}
\end{center}
\caption{\label{tab:ratios} Mean and angle of spacing ratio distributions obtained with $10^2$ samples of random  $10^4\times 10^4$ matrices rounded to the 4th digit. The matrix ensembles correspond to the ones shown in Figure~\ref{fig:spacing_ratios}.}
\end{table}

\section{discussion}\label{sec:discussion}
\subsection{Summary of results}
Motivated by the incapability of dense random Kolmogorov operators to capture spectral features of model Markov processes, we introduced and analyzed an ensemble of sparse random Kolmogorov operators. We showed that, if the number of non-zero elements per column (and row) $\varphi$ increases with the matrix size $D$,   the bulk of the spectrum is shifted away from the stationary eigenvalue $0$ in the limit of large matrix size $D$.  This is  independent of the weight distribution, i.e. of the distribution of the nonzero matrix elements. 

In contrast, the spectral edges depend on the tails of the weight distribution. The tails of the weight distribution determine, together with $\varphi$, the tails of diagonal elements of generator matrices. We numerically showed that the spectral edges are well approximated by the extremes of the diagonal elements. From extreme value theory it follows that for  diagonal distributions with power-law left tails (this includes among others edge weights being uniform, exponential, $\chi^2$, gamma or beta distributed), the average spectral gap decreases as a power-law in $D$ for fixed $\varphi$, is constant for $\varphi\sim \log D$ and increases, whenever $\varphi$ increases with $D$ substantially faster than $\log D$. 

A similar approach was used to calculate the horizontal extent, given by the eigenvalue with the smallest real part. We linked the horizontal extent to the largest diagonal element (in magnitude) of the generator matrix and used extreme value theory to calculate the latter. 

Finally, we showed that complex spacing ratio distributions of generator matrices with $\varphi\ge 2$ follow the distribution typical of Ginibre's Orthogonal Ensemble, while there is a strong anomaly for $\varphi=1$.

\subsection{Open questions}

(1) We have introduced sparsity to model $\mathcal{K}$-generators of physical Markov processes, and have used the sparsity to tune spectral features of the generators.  There are other ways of providing random matrices with a structure that models physical constraints (e.g., locality).  E.g., one could   consider banded matrices \cite{Casati_etal_PRL1990banded_matrices, Casati_et_al_JourPhysA1991banded_matrices, Fyodorov_Mirlin_PRL1991RMT_banded, Fyodorov_Mirlin_PRB1995banded_matrices, Sodin_AnnMath2010_banded_matrices,Erdos_Knowles_CommMathPhys2011banded_matrix, Erdos_et_al_CommMathPhys2013_banded_matrices, Erdos_et_al_AnnHenriPoinc2013banded_matrices, Spencer_OxfordHandbookRMT2015, Bourgade_Review2018bandedRMT} or matrices with decaying off-diagonal terms \cite{Mirlin_et_al_PRE1996power_law_matrices, Fyodorov_et_al_JourStatMech2009_ultra_metric_matr, Spencer_OxfordHandbookRMT2015} or temperature based models \cite{Mosam_Vidaurre_Giuli_PRE2021temp_markov_model}.  These are alternate routes to tuning spectral features.  To the best of our knowledge, generators of CTMCs with such structures have not yet been considered.

(2) The application of extreme value theory to find the limiting distribution of the spectral edges relied on the observation that the spectral edges are well approximated by the minimum and the maximum of the diagonal of the generator matrix. By the Courant-Fisher theorem, the extremes of the diagonal are upper and lower bounds, respectively, for symmetric generators. In this case, a concentration of the largest eigenvalue in magnitude around the maximum of the diagonal was shown in \cite{Bandeira_FoundCompMath2018}. An analytical treatment of general non-symmetric generators and the spectral gap is to the best of our knowledge not known. We hope that our results motivate a rigorous investigation of the connection between the spectral edges and the diagonal of the generator matrix.

(3) Generators of CTMCs have real entries and thus their eigenvalues are real or come in complex conjugate pairs. In the investigation of correlations between eigenvalues, we left out real eigenvalues.  
The appearance of a large number of real eigenvalues in the spectrum of non-Hermitian matrices is a phenomenon of wide interest \cite{Lehmann_Sommers_PRL1991random_real_matrix, Edelman_et_al_JourAmMathSoc1994real_eigenvalues, Edelman_JourMulAna1997real_eigenvalues, Kanzieper_Akemann_PRL2005real_eigenvalues, Forrester_Nagao_PRL2007real_gin_ensemble, Sommers_Wieczorek_JourPhysA2008real_ginibre, Timm_PRE2009, Khoruzhenko_Sommers_Zyckowski_PRE2010rand_orth_matrices, Tao_Vu_AnnProb2015, Tarnowski_PRE2022real_eigvals, Byun_et_al_IntMathResNot2023real_eigvals_elliptic}. 
For real Ginibre matrices, the average number of real eigenvalues is $\sim D^{-1/2}$ \cite{Edelman_et_al_JourAmMathSoc1994real_eigenvalues, Edelman_JourMulAna1997real_eigenvalues, Kanzieper_Akemann_PRL2005real_eigenvalues} while for dense generators of CTMCs, it is substantially larger \cite{Timm_PRE2009}. We observed that the fraction of real eigenvalues is larger for small $\varphi$ and smaller for larger $\varphi$ (not presented).  Understanding of the functional dependence of the number of real eigenvalues for sparse CTMC generators is an interesting problem.

(4) We focused on the location and extent of the bulk spectrum as well as the spectral edges.   One could inquire about the evolution of other features of the spectral distribution as a function of sparsity, e.g., about the envelope of the spectral distribution.   In \cite{Tarnowski_Denisov_etal_PRE2021} the spectral density of dense random CTMCs was described by the convolution of two asymptotically free matrices, leading to the prominent spindle shape.  Free probability arguments break down for sparse random CTMCs. Analytical tools which have been employed to calculate the spectral density of sparse, random matrices include replica tricks \cite{Bray_Rodgers_PRB1988sparse_ctmc, Rodgers_Bray_PRB1988sparse_ctmc, Cavagna_Giardina_Parisi_PRL1999sparse_laplacian, Kuehn_JourPhysA2008sparse_mc, Susca_Vivo_Kuehn_SciPost2021replica_cavity}, single defect and effective medium approximations  \cite{Biroli_Monasson_JourPhysA1999_sda, Semerjian_Cugliandolo2002sparse_sda_ema, Dorogovtsev_et_al_PRE2003EMA}, supersymmetry-based techniques \cite{Mirlin_Fyodorov_JourPhysA1991supersymmetry, Pipattana_Evnin_2022supersymmetry} and the cavity approach \cite{Rogers_Castillo_Kuehn_PRE2008cavity_symmetric, Rogers_Castillo_PRE2009cavity_method_nonhermitian, Susca_Kuehn_JourPhysA2019largest_eigenvalue_sparse_mc, Susca_Vivo_Kuehn_SciPost2021replica_cavity}. Spectral properties of symmetric, sparse, random CTMCs have been investigated with the cavity method \cite{Margiotta_Kuehn_Sollich_IOP2019sparse_random_symmetric_ctmc, Tapias_Parpotzki_Sollich_JourStatMech2020Barrat_Mezard, Tapias_Sollich_PRE2022Barrat_Mezard} and with supersymmetric approaches \cite{Pipattana_Evnin_2022supersymmetry}. Investigations of the spectral density of non-symmetric sparse, random Kolmogorov operators with the above  methods might be an interesting objective.

(5) We have considered sparse generators of CTMCs based on strongly connected, sparse random graphs. It is an open question whether our results can be generalized to other sparse graph ensembles. One potential avenue to explore are directed Erdös-Renyi (dER) graphs.

In dER  graphs, the probability of an edge connecting any two vertices is $0<p\le 1$. For a dER graph to be strongly connected with a high probability, the value of $p$ must exceed $\sim\log D/D$ \cite{Palasti_StudSciMathHun1966dER, Graham_Pike_Eucl2008ER_strong_conn}. As a result, the average degree of the vertices must increase logarithmically with $D$ to ensure strong connectivity. Consequently, the range of constant average vertex degree and increasing vertex number $D$ is excluded.

Nonetheless, modifying the dER graph by enforcing a minimum (in- and out-) degree $\ge 2$ guarantees strong connectivity with high probability \cite{Pittel_2018}. Exploring the spectral properties of CTMC generators based on dER graphs may represent a promising next step towards generalizing our results.

(6) Finally, there is an interesting question: What could 'sparsity' mean in the quantum limit? Namely, what is 'sparsity' for Lindblad operators? 

Here we start from the genetic link which allows to obtain a generator of a classical (quantum) Markovian evolution as a properly normalized non-probability (trace) preserving stochastic map (channel). Eq.~(3), where $\cal{M}$ is a no-probability-preserving map and $\cal{J}$ takes care of normalization, illustrates this link in the case of Kolmogorov operators. It seems to be  intuitive that sparsity $\varphi$ of a stochastic matrix  can be associated with  rank $r$ of a quantum channel~\cite{Watrous_2018quantinf}.

Thus, in the ultimate limit $\varphi=r=1$,  quantum versions of stochastic maps --  
that are permutations now -- are rank-one  channels -- that are unitaries.  It is tempting to extend this analogy beyond the limit $\varphi=r=1$ and state that mixed-unitary channels (convex combinations of unitaries) are quantum versions of  bistochastic matrices (that are, according to Birkhoff, convex combinations of permutations~\cite{Birkhoff_1946}). 
    
However, there is also notion of double stochastic (or "unital") channels~\cite{Watrous_2018quantinf}. The two classes, mixed unitaries and bistochastic channels, are not identical: there are double stochastic channels  that are  outside of the convex hull of unitaries~\cite{Landau_Streater_LinAlg1993birkhoff_thm}. What class to associate with classical bistochastic maps is then 'a matter of taste'~ \cite{Zyczkowski_pers_comm}. To resolve the dichotomy, one could rely on the concept of super-decoherence~\cite{Tarnowski_Denisov_etal_PRE2021} and state that all channels which have classical bistochastic matrices as their 
fully decohered versions, are quantum analogues of the matrices. In this case the broader class of double stochastic channels is chosen~\cite{Zyczkowski_pers_comm}. 

The superdecoherence-based reasoning can also be applied to generators. In this case the unitary (Hamiltonian) part of a Lindblad operator does not play any role since it vanishes in the limit of complete decoherence~\cite{Tarnowski_Denisov_etal_PRE2021} and the quantum 'sparsity' is defined by the rank of the dissipative part (the minimal number of jump operators).

Interestingly, Lindblad operators of ultra-low rank $r=1$ were considered in Refs.~\cite{Can_JourPhysA2019randLindblad} and \cite{Can_et_al_PRL2019Lindblad_gaps}.
Features similar to ones we detected for ultra-sparse Kolmogorov operators were reported (e.g., the spectral gap is defined by a real-valued  outlier with  position independent of $D$). 

\begin{acknowledgments}
The authors thank Gernot Akemann, Alexander~van~Werden, Ioana~Dumitriu,  Toma\u{z}~Prosen, Karol \.{Z}yczkowski, and Wojciech~Tarnowski for helpful discussions and comments.
This research is supported by the Deutsche Forschungsgemeinschaft through SFB 1143 (project-id 247310070) [GN and MH] and Research Council of Norway,
project IKTPLUSS-IKT og digital innovasjon - 333979 (as a part of the ERA-NET project "DQUANT: A Dissipative Quantum Chaos perspective on Near-Term Quantum Computing") [SD].
\end{acknowledgments}

\appendix

\section{Models in Figure 1}\label{sec:appendix_systems_description}
In the following, we denote spin creation and annihilation operators as $\sigma^+$ and $\sigma^-$, respectively. We split the generator matrix $\mathcal{K}$ into an off-diagonal matrix $\mathcal{M}$ and a diagonal matrix $\mathcal{J}$ such that $\mathcal{K}=\mathcal{M}-\mathcal{J}$ and the diagonal entries of $\mathcal{J}$ are the sums of the columns of $\mathcal{M}$.

In Figure~\ref{fig:spectra_systems} (b) we show the TASEP on a ring with $L=12$ sites and staggered hopping amplitudes. The $\mathcal{M}$ matrix is given by \cite{Sa_Ribeiro_Prosen_PRX2020}
\begin{equation}
	\mathcal{M} = \frac{1}{2}\sigma_1^+ + \frac{1}{2} \sigma_1^- + 
	\sum_{j=1}^L p_j \sigma_{j}^- \sigma_{j+1}^+ ,
\end{equation}
where $p_j = 1$ if $j$ is even and $p_j=0.2$ if $j$ is odd.

In Figure~\ref{fig:spectra_systems} (c) we show the ASEP on a chain of length $L=12$ with open boundary conditions and next nearest neighbor hopping. The $\mathcal{M}$ matrix is given by
\begin{equation}
	\mathcal{M} = \sigma_1^+ + \sigma_L ^- + \sum_{j=1}^L \sigma_{j}^- \sigma_{j+1}^+ + \sum_{j=1}^{L/2}  \sigma_{2j}^- \sigma_{2j+2}^+.
\end{equation}

In Figure~\ref{fig:spectra_systems} (d) we show the spectrum of a single particle hopping on a $65\times 65$ grid with  periodic boundary conditions and random hopping amplitudes. The $\mathcal{M}$ matrix is given by
\begin{equation}
	\mathcal{M} = \sum_{ \langle (i,j), (i', j') \rangle} p_{(i,j)\to (i',j')} \sigma_{i,j}^- \sigma_{i',j'}^+
\end{equation}
where $\langle\dots\rangle$ denotes summation over nearest neighbors and $p_{(i,j)\to (i',j')}$ are randomly uniformly chosen between 0 and 1 under the constraint that $p_{(i,j)\to (i',j')}  = 1 - p_{(i',j')\to (i,j)}$.  This diffusion model can of course be extended to many particles, but we choose to show the single-particle sector here.

In Figure~\ref{fig:spectra_systems} (e) we show the spectrum of a contact process \cite{Harris_AnnProb1974} on a chain with $L=12$ sites and open boundary conditions. The master equation is generated by $-H$, where $H$ is given by
\begin{equation}
	H = \sum_{i=1}^L M_i + \sum_{i=1}^{L-1} \left[ n_i Q_{i+1} + Q_i n_{i+1} \right],
\end{equation}
and
\begin{equation}
	M = \begin{pmatrix}
		0 &-1 \\ 0 & 1
	\end{pmatrix},
	\quad
	n = \begin{pmatrix}
		0 & 0 \\ 0 & 1
	\end{pmatrix},
	\quad
	Q = \begin{pmatrix}
		1 & 0 \\ -1 & 0
	\end{pmatrix}.
\end{equation}

Finally, in Figure~\ref{fig:spectra_systems} (f) we show the spectrum of the generator matrix $\mathcal{K}$ of a gene transcription model taken from \cite{Sevier_etal_PNAS2016_transcrpt_noise}. The following master equations model the accumulation and release of mechanical strain of DNA during transcription. The parameters chosen for the spectral data in Figure~\ref{fig:spectra_systems} (f) are the mRNA transcription rate $r=2$ and decay rate $\lambda=0.05$, the maximum number of transcripts until no further strain can be put on the DNA $m_c=10$, the relaxation rate of the DNA string $g=0.05$ and a maximum number of transcription events $m_{\max}=400$ to make the generator matrix $M$ finite. By $m$ we denote the number of current transcripts and by $\alpha$ the number of transcripts made since the last relaxation event. Then for $0\le m \le m_{\max}$ and $1\le\alpha\le m_c-1$ the master equation reads
\begin{align}
	\frac{d}{dt} P_\alpha
	&= -(r+g+\lambda m) P_\alpha(m,t) + \lambda(m+1) P_\alpha(m+1,t) \nonumber \\
	&+ rP_{\alpha-1}(m-1,t)
\end{align}
while for $\alpha=0$ we have
\begin{align}
	\frac{d}{dt} P_0 
	&= -(r+g+\lambda m) P_0(m,t) + \lambda(m+1) P_0(m+1,t) \nonumber\\
	&+ g\sum_{\alpha=0}^{m_c} P_\alpha(m,t)
\end{align}
and for $\alpha=m_c$
\begin{align}
	\frac{d}{dt} P_{m_c} 
	&= -(g+\lambda m) P_{m_c}(m,t) + \lambda(m+1)P_{m_c}(m+1,t) \nonumber \\
	&+rP_{m_c-1}(m-1,t).
\end{align}

\section{Sampling of sparse random generators}

In order to obtain a sparse random generator matrix, our approach involves first sampling a random directed graph with $D$ vertices and both in- and out-vertex degrees of $\varphi$. Subsequently, the non-zero elements of the corresponding adjacency matrix are sampled from a common positive distribution. This procedure results in the off-diagonal matrix $\mathcal{M}$. The random Markov generator matrix is then constructed as $\mathcal{K}=\mathcal{M}-\mathcal{J}$, where $\mathcal{J}$ is a diagonal matrix with diagonal elements equal to the sums of the columns of $\mathcal{M}$.

The random directed graph is generated by iteratively connecting each vertex to $\varphi$ other vertices, while rejecting edges if the corresponding vertex already has $\varphi$ incoming edges. For the final vertices, it may not be feasible to connect to other vertices without violating the constraint of $\varphi$ incoming edges for each vertex. In such cases, the entire process is restarted. To mitigate the risk of restarting the procedure, we reduce the probability of connecting to a vertex that already has a high degree. Following this approach, we find that we rarely need to restart the algorithm for the matrix sizes and vertex degrees $\varphi$ examined in this study.

To compute the eigenvalues of the Markov matrices, we utilize an exact diagonalization method, while the Arnoldi method is employed to calculate the spectral gap. We deem an eigenvalue to have converged once the norm of the residuals of the Schur vectors is less than $10^{-12}$.

\section{Analytical results for the bulk spectrum}\label{sec:appendix_bulk}

In this section, we will derive the analytical results of the estimated mean $\mu(\lambda)$ in Eq.~\eqref{eq:est_mean} and the estimated pseudo-variance in Eq.~\eqref{eq:sigma2_analytical} in the main text and show that $\frac{1}{D} \sum_{j=1} \lambda_j$ concentrates around its average $\langle\dots\rangle$.

Denote by $\iota$ the function $\iota:\{1,\dots,\varphi\}\times\{1,\dots,D\}\to\{1,\dots,D\}^2$ with $\iota(l,j)=(i,j)$ where $i$ is the $l$th non-zero index in column $j$ in $\mathcal{M}$. Note that $\iota(l,j)=(i,j)$ implies $i\neq j$ and $l\to\iota(l,j)$ is injective for fixed $j$. Further, let in this appendix the location of the bulk be denoted as
\begin{equation*}
	\mu(\lambda) = \frac{1}{D} \sum_{j=1}^D \lambda_j = \frac{1}{D} \tr(\mathcal{K}).
\end{equation*}
and the pseudo-variance as
\begin{align}\label{eq:app_sigma}
	\sigma^2(\lambda) 
	= \frac{1}{D}\sum_{j=1}^D \lambda_j^2 - \left( \frac{1}{D}\sum_{j=1}^D \lambda_j \right)^2 \nonumber \\
	= \frac{\tr(\mathcal{K}^2)}{D} - \frac{\tr(\mathcal{K})^2}{D^2}.
\end{align}
Here we explicitly do not include the averaging over the matrix ensemble $\langle\dots\rangle$ in contrast to the main text.

\subsection{Location}
The average value with respect to $\langle\dots\rangle$ of the location $\mu(\lambda)$ can then be computed as
\begin{align*}
	\llangle\mu(\lambda)\rrangle
	&=\llangle \frac{1}{D} \tr(\mathcal{K}) \rrangle
	= \frac{1}{D} \sum_{j=1}^D \llangle K_{jj} \rrangle \\
	&= \frac{1}{D} \sum_{j=1}^D \sum_{l=1}^\varphi \llangle K_{\iota(l, j)} \rrangle
	= - \varphi \mu_0,
\end{align*}
where we used that $\llangle K_{\iota(l, j)} \rrangle=-\mu_0$. This is Eq.~\eqref{eq:est_mean} in the main text. Similar,
\begin{align*}
	\llangle \tr(\mathcal{K})^2 \rrangle
	&=\sum_{j_1,j_2=1}^D \sum_{l_1,l_2=1}^\varphi \llangle K_{\iota(l_1,j_1)} K_{\iota(l_2,j_2)} \rrangle \\
	&=\sum_{j=1}^D \left[ \sum_{l=1}^\varphi \llangle K_{\iota(l,j)}^2 \rrangle 
	+ \sum_{l_1\neq l_2} \llangle K_{\iota(l_1,j)} K_{\iota(l_2,j)} \rrangle \right] \\
	&+ \sum_{j_1\neq j_2} \sum_{l_1,l_2=1}^\varphi  \llangle K_{\iota(l_1,j_1)} K_{\iota(l_2,j_2)} \rrangle.
\end{align*}
Although the off-diagonal elements of $\mathcal{K}$ are weakly dependent because of the constraint that the number of non-zero elements per row and column has to equal $\varphi$, the non-zero elements $K_{\iota(l,j)}$ are independent. Hence, $\llangle K_{\iota(l_1,j)} K_{\iota(l_2,j)} \rrangle = \llangle K_{\iota(l_1,j)} \rrangle \llangle K_{\iota(l_2,j)} \rrangle$ and $\llangle K_{\iota(l_1,j_1)} \rrangle \llangle K_{\iota(l_2,j_2)} \rrangle$, so
\begin{align*}
	\llangle \tr(\mathcal{K})^2 \rrangle
	&= D\varphi(\sigma_0^2+\mu_0^2) + D\varphi(\varphi-1)\mu_0^2 + D(D-1) \varphi^2 \mu_0^2 \\
	&= D\varphi \sigma_0^2 + (D\varphi\mu_0)^2,
\end{align*}
where we used that the second moment $\llangle K_{\iota(l,j)}^2 \rrangle$ equals $\sigma_0^2+\mu_0^2$. This implies that
\begin{align*}
	\llangle\mu(\lambda)^2\rrangle - \langle\mu\rangle^2
	= \llangle \frac{\tr(\mathcal{K})^2}{D^2} \rrangle - \llangle \frac{\tr(\mathcal{K})}{D} \rrangle^2
	= \frac{\varphi\sigma_0^2}{D}.
\end{align*}
The right-hand side vanishes for increasing $D$ and $\varphi$ growing slower with $D$ than linear. Relatively to $\llangle \mu(\lambda) \rrangle$ the typical deviation of $\mu(\lambda)$ from its average value always vanishes for either increasing $D$ or $\varphi$, as
\begin{align*}
	\frac{\sqrt{\llangle\mu(\lambda)^2\rrangle - \langle\mu\rangle^2}}{|\llangle\mu(\lambda)\rrangle|}
	= \frac{\sigma_0}{\mu_0} \left(\varphi D\right)^{-1/2}.
\end{align*}

\subsection{Complex pseudo-variance}

The first term in the averaged pseudo-variance given by Eq.~\eqref{eq:app_sigma} can be calculated as
\begin{align}\label{eq:app_sigma_tr_M2}
	\left\langle \tr(\mathcal{K}^2) \right\rangle
	&= \sum_{i,j=1}^D \langle K_{ij} K_{ji} \rangle \nonumber \\
	&= \sum_{i=1}^D \langle K_{ii}^2 \rangle + \sum_{i\neq j} \langle K_{ij} K_{ji} \rangle.
\end{align}
We proceed with $\sum_{i=1}^D \langle K_{ii}^2 \rangle$ in Eq.~\eqref{eq:app_sigma_tr_M2} and get
\begin{align}\label{eq:app_sigma_Mii}
	\sum_{i=1}^D \left\langle K_{ii}^2 \right\rangle \nonumber
	&= \sum_{i=1}^D \left\langle \left( - \sum_{j\neq i} K_{ji} \right)^2 \right\rangle \nonumber\\
	&= \sum_{i=1}^D \sum_{j,l\neq i} \left\langle K_{ji} K_{li} \right\rangle \nonumber\\
	&= \sum_{i=1}^D \sum_{j\neq i} \langle K_{ji}^2 \rangle 
	+ \sum_{i=1}^D \sum_{j,l\neq i; j\neq l} \left\langle K_{ji}\rangle \langle K_{li} \right\rangle.
\end{align}
The former sum in Eq.~\eqref{eq:app_sigma_Mii} is given by
\begin{align}\label{eq:app_sigma_Mii1}
	\sum_{i=1}^D \sum_{j\neq i} \langle K_{ji}^2 \rangle
	&= \sum_{i=1}^D \sum_{l=1}^\varphi \langle K_{\iota(l,i)}^2 \rangle
	= D\varphi (\sigma_0^2 +\mu_0^2),
\end{align}
where again we used that $\langle K_{\iota(l,i)}^2 \rangle = \sigma_0^2 + \mu_0^2$, while the latter sum in Eq.~\eqref{eq:app_sigma_Mii} is
\begin{align}\label{eq:app_sigma_Mii2}
	%\sum_{i=1}^D \sum_{j,l\neq i; j\neq l} \left\langle M_{ji} M_{li} \right\rangle
	%&= \sum_{i=1}^D \sum_{j\neq i} \sum_{l\neq i,j} \langle M_{ji} M_{li} \rangle
	\sum_{i=1}^D &\sum_{j,l\neq i; j\neq l} \left\langle K_{ji}\rangle \langle K_{li} \right\rangle \nonumber\\
	&= \sum_{i=1}^D \sum_{k=1}^\varphi \sum_{n=1; \iota(n,i)\neq\iota(k,i)}^\varphi \langle K_{\iota(k,i)} \rangle \langle K_{\iota(n,i)} \rangle \nonumber \\
	&=D\varphi(\varphi-1) \mu_0^2.
\end{align}
Combining Eq.~\eqref{eq:app_sigma_Mii1} and Eq.~\eqref{eq:app_sigma_Mii2} we get
\begin{align*}
	\sum_{i=1}^D \left\langle K_{ii}^2 \right\rangle 
	&= D \varphi (\sigma_0^2 + \mu_0^2) + D\varphi(\varphi-1) \mu_0^2\\
	&= D \varphi \sigma_0^2 + D \varphi^2 \mu_0^2.
\end{align*}
Now, we are left with calculating $\sum_{i\neq j} \langle K_{ij}K_{ji} \rangle$, the second term in Eq.~\eqref{eq:app_sigma_tr_M2},
\begin{align*}
	\sum_{i\neq j} \langle K_{ij} K_{ji} \rangle
	&= \sum_{i=1}^D \sum_{l=1}^\varphi \llangle K_{\overline{\iota(l,i)}} M_{\iota(l,i)} \rrangle,
\end{align*}
where the $\overline{\iota}$ denotes swapping the first and second component. Note that $K_{\overline{\iota(l,i)}}$ is not necessarily a non-zero entry of $\mathcal{K}$, hence $K_{\overline{\iota(l,i)}}$ and $K_{\iota(l,i)}$ depend weakly on each other. In the large $D$ limit we can assume that the dependence is sufficiently weak and we treat $K_{\overline{\iota(l,i)}}$ and $K_{\iota(l,i)}$ as independent, thus $\llangle K_{\overline{\iota(l,i)}} K_{\iota(l,i)} \rrangle = \mu_0 \llangle K_{\overline{\iota(l,i)}}\rrangle$. By the assumed independence the mean of \textit{every} entry in the $i$th row, except the diagonal, is $\llangle K_{\overline{\iota(l,i)}}\rrangle = \frac{\varphi}{D} \mu_0$. Hence,
\begin{align*}
	\sum_{i\neq j} \langle K_{ij} K_{ji} \rangle
	&= \sum_{i=1}^D  \frac{1}{D} \varphi^2 \mu_0^2 
	= \varphi^2 \mu_0^2.
\end{align*}

Collecting the above results we arrive at
\begin{align*}
	\left\langle \tr(\mathcal{K}^2) \right\rangle
	&=  D \varphi \sigma_0^2 + D \varphi^2 \mu_0^2 + \varphi^2 \mu_0^2 \\
	&= D \varphi \sigma_0^2 + (D+1) \varphi^2 \mu_0^.
\end{align*}

The second term of the averaged pseudo-variance in Eq.~\eqref{eq:app_sigma} has been calculated in the previous subsection,
\begin{align*}
	\llangle \tr(\mathcal{K})^2 \rrangle =  D\varphi \sigma_0^2 + (D\varphi\mu_0)^2
\end{align*}

Finally, we can evaluate
\begin{align*}
	\llangle \sigma^2(\lambda) \rrangle
	&= \llangle \frac{\tr(\mathcal{K}^2)}{D} \rrangle - \llangle \frac{\tr(\mathcal{K})^2}{D^2} \rrangle\\
	&= \varphi \sigma_0^2 + \varphi^2 \mu_0^2 + \frac{1}{D}\varphi^2 \mu_0^2 - \frac{1}{D}\varphi \sigma_0^2 - \varphi^2\mu_0^2\\
	&= \varphi \left( \sigma_0^2 + \frac{\varphi}{D} \mu_0^2 - \frac{1}{D}\sigma_0^2 \right),
\end{align*}
which is Eq.~\eqref{eq:sigma2_analytical} in the main text.

\section{Bound of spectral gap for symmetric $M$}\label{sec:appendix_gap}

In this section, we give the proof of Eq.~\eqref{eq:gap_ineq_min}. Let $\mathcal{K}=\mathcal{M}-\mathcal{J}$ be a symmetric generator matrix. By Eq.~\eqref{eq:ritz_gap} we have to show that $v^t \mathcal{K} v \le \min_{1\le l \le D} J_{ll} + O\left(D^{-1}\right)$ for the vector $v$ given 
\begin{equation*}
	v_i = \begin{cases}
		\sqrt{ 1 - \frac{1}{D}} &i=l \\
		-\frac{1}{\sqrt{D(D-1)}} &i\neq l,
	\end{cases}
\end{equation*} 
where $1\le l\le D$ is arbitrary. It is easy to see that $|v|=1$ and $v\perp v_1$. So we proceed with
\begin{align}\label{eq:derivation_upper_bound}
	\gamma_*
	&\le v^t (\mathcal{J}-\mathcal{M}) v
	= \sum_{i,j=1}^D v_i v_j (\mathcal{J}-\mathcal{M})_{ij} \nonumber \\
	&= \sum_{i=1}^D v_j^2 J_{jj} - \sum_{i,j=1}^D v_i v_j M_{ij}  \nonumber \\
	&= \sum_{i,j=1}^D v_j^2 M_{ij} - \sum_{i,j=1}^D v_iv_j M_{ij} \nonumber\\
	&= \sum_{i,j=1}^D v_j M_{ij} (v_j-v_i).
\end{align}
Note that any summand in Eq.~\eqref{eq:derivation_upper_bound} where either $i=j=l$ or $i\neq l$ and $j\neq l$ is zero. Inserting the definition of $v$ we get
\begin{align}\label{eq:derivation_upper_bound2}
	\gamma_* 
	&\le \sum_{i\neq l} v_l M_{il} (v_l - v_i) + \sum_{j\neq l} v_j M_{lj} (v_j - v_l) \nonumber \\
	&= \sum_{i\neq l} \sqrt{1-\frac{1}{D}} M_{il} \left( \sqrt{1 - \frac{1}{D}} + \frac{1}{\sqrt{D(D-1)}}\right) \nonumber \\ 
	&- \sum_{j\neq l} \frac{1}{\sqrt{D(D-1)}} M_{lj} \left( - \frac{1}{\sqrt{D(D-1)}} - \sqrt{1-\frac{1}{D}}\right) \nonumber \\
	&=\left( \sqrt{1-\frac{1}{D}} + \frac{1}{\sqrt{D(D-1)}}\right) \nonumber \\
	&\times \sum_{i\neq l}  \left[ \sqrt{1-\frac{1}{D}} M_{il} + \frac{1}{\sqrt{D(D-1)}} M_{li}\right].
\end{align}
After collecting all the prefactors in Eq.~\eqref{eq:derivation_upper_bound2} the spectral gap is upper-bounded by
\begin{equation*}
	\gamma_* 
	\le \sum_{i\neq l} \left[ M_{il} + \frac{1}{D-1} M_{li} \right]
	= J_{ll}+\frac{1}{D-1}\tilde{J}_{ll},
\end{equation*}
where we denote $\tilde{J}_{ll}=\sum_{i\neq l} M_{il}$. As the number of non-zero elements of $\mathcal{M}$ in every row and column is the same, the distribution of $J_{ll}$ and $\tilde{J}_{ll}$ coincide. In the limit of large $D$, $J_{ll}$ and $\tilde{J}_{ll}$ are independent. Thus we can approximate $\gamma_* \le J_{ll} + O\left(D^{-1}\right)$ at least for $\varphi\ll D$. As the index $l$ was chosen arbitrarily we get
\begin{equation*}\label{eq:app_gap_J}
	\gamma_* \le \min_{1\le l \le D} J_{ll} + O\left(D^{-1}\right),
\end{equation*}
which is Eq.~\eqref{eq:gap_ineq_min} in the main text.

\bibliography{lit_sparse}

\end{document}